\documentclass[numberedappendix,preprint2]{emulateapj} 
\usepackage{amsmath,natbib,graphicx}
\usepackage{epsf}
\input{epsf}
\usepackage{epstopdf}
\usepackage{multirow}
\usepackage{subfigure}
\usepackage{epsfig}
\usepackage{color}
\usepackage{lscape}
\usepackage{appendix}

\usepackage{ifthen}

\DeclareGraphicsExtensions{.jpg,.pdf,.png,.eps,.ps}
\graphicspath{{figures/}}

\newcommand{\degs}{deg$^2$}

\newcommand{\degss}{deg$^2$ }
\newcommand{\fivehs}{\textsc{ra5h30dec-55} }
\newcommand{\twentythreehs}{\textsc{ra23h30dec-55} }
\newcommand{\fsz}{\ensuremath{f_\mathrm{\tiny{SZ}}}}
\def\arcsec{\hbox{$^{\prime\prime}$}}
\newcommand{\noisenntysurvey}{42}
\newcommand{\noiseonftysurvey}{18}
\newcommand{\noisettwntysurvey}{85}

\newcommand{\noiseonftybeamcal}{14.8}
\newcommand{\noisettwntybeamcal}{36.1}
\newcommand{\gaussratonfty}{0.86}
\newcommand{\gaussratttwnty}{0.87}
\newcommand{\noiseonftygauss}{17}
\newcommand{\noisettwntygauss}{41}
\newcommand{\noiseonftydecon}{19.7}
\newcommand{\noisettwntydecon}{46.9}
\newcommand{\noiseonftydecongauss}{18}
\newcommand{\noisettwntydecongauss}{43}
\newcommand{\netonftylow}{380}
\newcommand{\netonftyhigh}{540}
\newcommand{\netttwntylow}{640}
\newcommand{\netttwntyhigh}{850}
\newcommand{\netonftyarray}{29}
\newcommand{\netttwntyarray}{74}
\newcommand{\noiseonftypredraw}{11}
\newcommand{\noisettwntypredraw}{29}
\newcommand{\noiseonftypredeff}{13.4}
\newcommand{\noisettwntypredeff}{34.7}

\begin{document}

\title{The First  Public Release of South Pole Telescope Data: Maps of a 95-square-degree Field 
from 2008 Observations}

\slugcomment{Published in the Astrophysical Journal, 743, 90}

\author{
 K.~K.~Schaffer,\altaffilmark{1,2,3}
 T.~M.~Crawford,\altaffilmark{1,4}
 K.~A.~Aird,\altaffilmark{5}
 B.~A.~Benson,\altaffilmark{1,2}
 L.~E.~Bleem,\altaffilmark{1,6}
 J.~E.~Carlstrom,\altaffilmark{1,2,4,6,7}
 C.~L.~Chang,\altaffilmark{1,2,7}
 H.~M. Cho,\altaffilmark{8}
 A.~T.~Crites,\altaffilmark{1,4}
 T.~de~Haan,\altaffilmark{9}
 M.~A.~Dobbs,\altaffilmark{9}
 E.~M.~George,\altaffilmark{10}
 N.~W.~Halverson,\altaffilmark{11}
 G.~P.~Holder,\altaffilmark{9}
 W.~L.~Holzapfel,\altaffilmark{10}
 S.~Hoover,\altaffilmark{1,6}
 J.~D.~Hrubes,\altaffilmark{5}
 M.~Joy,\altaffilmark{12}
 R.~Keisler,\altaffilmark{1,5}
 L.~Knox,\altaffilmark{13}
 A.~T.~Lee,\altaffilmark{10,14}
 E.~M.~Leitch,\altaffilmark{1,4}
 M.~Lueker,\altaffilmark{15}
 D.~Luong-Van,\altaffilmark{5}
 J.~J.~McMahon,\altaffilmark{16}
 J.~Mehl,\altaffilmark{1}
 S.~S.~Meyer,\altaffilmark{1,2,4,6}
 J.~J.~Mohr,\altaffilmark{17,18,19}
 T.~E.~Montroy,\altaffilmark{20}
 S.~Padin,\altaffilmark{1,4,15}
 T.~Plagge,\altaffilmark{1,4}
 C.~Pryke,\altaffilmark{1,2,4,21}
 C.~L.~Reichardt,\altaffilmark{10}
 J.~E.~Ruhl,\altaffilmark{20}
 E.~Shirokoff,\altaffilmark{10}
 H.~G.~Spieler,\altaffilmark{14}
 B. Stalder,\altaffilmark{22}
 Z.~Staniszewski,\altaffilmark{15}
 A.~A.~Stark,\altaffilmark{22}
 K.~Story,\altaffilmark{1,6}
 K.~Vanderlinde,\altaffilmark{9}
 J.~D.~Vieira,\altaffilmark{15} and
 R.~Williamson,\altaffilmark{1,4}
}

\altaffiltext{1}{Kavli Institute for Cosmological Physics,
University of Chicago, 5640 South Ellis Avenue, Chicago, IL, USA 60637}
\altaffiltext{2}{Enrico Fermi Institute,
University of Chicago,
5640 South Ellis Avenue, Chicago, IL, USA 60637}
\altaffiltext{3}{Liberal Arts Department, 
School of the Art Institute of Chicago, 
112 S Michigan Ave, Chicago, IL, USA 60603}
\altaffiltext{4}{Department of Astronomy and Astrophysics,
University of Chicago,
5640 South Ellis Avenue, Chicago, IL, USA 60637}
\altaffiltext{5}{University of Chicago,
5640 South Ellis Avenue, Chicago, IL, USA 60637}
\altaffiltext{6}{Department of Physics,
University of Chicago,
5640 South Ellis Avenue, Chicago, IL, USA 60637}
\altaffiltext{7}{Argonne National Laboratory, 9700 S. Cass Avenue, Argonne, IL, USA 60439}
\altaffiltext{8}{NIST Quantum Devices Group, 325 Broadway Mailcode 817.03, Boulder, CO, USA 80305}
\altaffiltext{9}{Department of Physics,
McGill University, 3600 Rue University, 
Montreal, Quebec H3A 2T8, Canada}
\altaffiltext{10}{Department of Physics,
University of California, Berkeley, CA, USA 94720}
\altaffiltext{11}{Department of Astrophysical and Planetary Sciences and Department of Physics,
University of Colorado,
Boulder, CO, USA 80309}
\altaffiltext{12}{Department of Space Science, VP62,
NASA Marshall Space Flight Center,
Huntsville, AL, USA 35812}
\altaffiltext{13}{Department of Physics, 
University of California, One Shields Avenue, Davis, CA, USA 95616}
\altaffiltext{14}{Physics Division,
Lawrence Berkeley National Laboratory,
Berkeley, CA, USA 94720}
\altaffiltext{15}{California Institute of Technology, MS 249-17, 1216 E. California Blvd., Pasadena, CA, USA 91125}
\altaffiltext{16}{Department of Physics, University of Michigan, 450 Church Street, Ann  
Arbor, MI, USA 48109}
\altaffiltext{17}{Department of Physics,
Ludwig-Maximilians-Universit\"{a}t,
Scheinerstr.\ 1, 81679 M\"{u}nchen, Germany}
\altaffiltext{18}{Excellence Cluster Universe,
Boltzmannstr.\ 2, 85748 Garching, Germany}
\altaffiltext{19}{Max-Planck-Institut f\"{u}r extraterrestrische Physik,
Giessenbachstr.\ 85748 Garching, Germany}
\altaffiltext{20}{Physics Department, Center for Education and Research in Cosmology 
and Astrophysics, 
Case Western Reserve University,
Cleveland, OH, USA 44106}
\altaffiltext{21}{Department of Physics, University of Minnesota, 116 Church Street S.E. Minneapolis, MN, USA 55455}
\altaffiltext{22}{Harvard-Smithsonian Center for Astrophysics,
60 Garden Street, Cambridge, MA, USA 02138}

\shorttitle{South Pole Telescope Public Data Release}
\shortauthors{Schaffer et al.}

\email{kschaf2@saic.edu}

\begin{abstract}
The South Pole Telescope (SPT) has nearly completed a 2500~\degss 
survey of the southern sky in three frequency bands.  Here we 
present the first public release of SPT maps and associated data 
products.
We present arcminute-resolution maps at 150~GHz and 220~GHz of an approximately 95~\degss field centered at
R.A. $82.7^\circ$, 
decl.~$-55^\circ$.
The field
was observed to a depth of
approximately \noiseonftygauss~$\mu$K-arcmin
at 150~GHz and 
\noisettwntygauss~$\mu$K-arcmin at 
220~GHz during the 2008 austral winter season.  
Two variations on
map filtering and map projection are presented, one tailored for 
producing catalogs of galaxy clusters detected through their 
Sunyaev-Zel'dovich effect signature and one tailored for producing
catalogs of emissive sources.
We describe the 
data processing pipeline, and we present instrument response functions, 
filter transfer functions, and map noise properties.  All data products described in this paper
are available for download at http://pole.uchicago.edu/public/data/maps/ra5h30dec-55
and from the NASA Legacy Archive for Microwave Background Data Analysis
server.  This is the first step in the eventual release of data from the full 
2500 \degss SPT survey.
\end{abstract}


\section{Introduction}
\label{sec:intro}

\subsection{Signals in the Millimeter-wave Sky}
\label{sec:mmsky}
Millimeter-wavelength (mm-wave) maps of the sky contain rich
cosmological and astrophysical information. 
Away from the Galactic plane, the mm-wave sky is dominated by the cosmic
microwave background (CMB) at large and intermediate angular size scales.
The CMB features on scales of roughly five arcmin to many degrees primarily arise from
temperature fluctuations at the surface of last scattering, and
mm-wave measurements of these anisotropies have enabled powerful
constraints on cosmological models \citep[e.g.,][]{komatsu11,dunkley11,keisler11}.  
Interactions of CMB photons with matter can induce secondary anisotropies 
through processes such as gravitational lensing and scattering, and 
measurements of these additional anisotropies are potentially powerful probes 
of cosmic structure.  The CMB features on scales smaller than about five 
arcmin are dominated by secondary anisotropy due to the inverse 
Compton scattering of CMB photons by free electrons, known as the 
Sunyaev-Zel'dovich (SZ) effect \citep{sunyaev72}.  On these small scales, emissive 
extragalactic sources also contribute significantly to the mm-wave sky.
These sources are of considerable astrophysical and cosmological 
interest themselves, particularly the population of high-redshift, dusty, 
star-forming galaxies (DSFGs) that make up the bulk of the cosmic 
infrared background \citep[CIB,][]{lagache05}.
The current generation of mm-wave telescopes is just beginning to exploit 
the scientific potential of these small-scale signals.

The SZ effect consists of two components: the 
kinetic SZ (kSZ) effect
and the 
thermal SZ (tSZ) effect.
The kSZ effect is due to Doppler
shifting of CMB photons by the bulk velocity of electrons 
along the line of sight.  The tSZ effect is due to the
scattering of CMB photons by hot, thermally distributed electrons,
primarily in galaxy clusters.  This interaction
results in a spectral distortion of the CMB with 
a null at approximately
220~GHz.  At observing frequencies above this null, the tSZ
effect produces an increment in measured CMB flux, while at
frequencies below the null, it produces a decrement.
The resulting tSZ features in the mm-wave sky can be used to detect and
characterize massive galaxy clusters.
Galaxy clusters trace the largest peaks in the matter density field of
the universe, and their abundance as a function of mass and redshift
is a sensitive probe of structure growth.  Galaxy clusters
selected in a fine-angular-scale tSZ survey provide a nearly
mass-limited and nearly redshift-independent sample for constraining 
cosmological parameters such as the dark energy
equation of state parameter $w$ and the normalization of the matter power spectrum
$\sigma_8$ \citep[e.g.,][]{carlstrom02}.  Such constraints
are already being realized with just the first small fraction of data from 
large mm-wave surveys \citep{vanderlinde10,sehgal10}.  

Measurements of the tSZ power spectrum provide
additional cosmological constraints, independent of those 
from the primary CMB and from catalogs of individually detected clusters 
\citep{lueker10,dunkley11,shirokoff11}.  These measurements 
can also inform models of the physical processes in galaxy clusters
\citep[e.g.,][]{shaw10,battaglia10}.  Measurements of the kSZ 
power spectrum are sensitive to, among other processes, 
the reionization history of the universe \citep[e.g.,][]{zahn05}, and 
limits on kSZ from mm-wave measurements have already ruled out
some reionization scenarios \citep{mortonson10}.
 
Two populations of extragalactic
sources contribute most significantly to the mm-wave
sky away from the Galactic plane:  sources for which the flux decreases 
or is roughly constant with frequency, consistent with
synchrotron emission from active galactic nuclei (AGN); and sources
for which the flux increases with frequency, consistent with thermal
emission from DSFGs.  Since the 
discovery by the SCUBA instrument \citep{holland99} of a population of 
moderate-to-high-redshift DSFGs responsible for 
a significant fraction of the total CIB
emission, DSFGs have been an active area of mm-wave and 
sub-millimeter (sub-mm) research.  The recent discovery of 
a sub-population of strongly lensed, higher-redshift DSFGs
\citep{vieira10,negrello10} has further increased interest in these 
sources.  Measurements of synchrotron-emitting sources in mm-wave
bands have the potential to constrain models of AGN physics 
\citep[e.g.,][]{dezotti10}.  

\subsection{The South Pole Telescope and Survey}
\label{sec:spt}
The South Pole Telescope (SPT) is a 10-meter telescope designed to
survey a large area of the sky at mm and sub-mm wavelengths with
arcminute angular resolution and low noise \citep{ruhl04, padin08,
carlstrom11}.  The current SPT receiver is a three-band (95, 150, and 220~GHz)
bolometer camera optimized for studying the CMB and the tSZ effect.
Since the SPT was commissioned
in 2007, the majority of observing time has been spent on a survey of 2500 \degss
designated the SPT-SZ survey. 
The final data set from the SPT-SZ survey will consist
of maps for 19 contiguous subfields of
70 to 230~\degs, observed during the austral winter seasons
of 2008 through 2011.  The final depth 
for most of the survey will be approximately 
\noisenntysurvey, \noiseonftysurvey,  and 
\noisettwntysurvey~$\mu$K-arcmin\footnote{Throughout this work,
map signal and noise amplitudes are expressed in units of 
K-CMB, expressing deviations from the average measured intensity as
equivalent temperature fluctuations in the CMB.} 
at 95, 150, and 220~GHz, with 
roughly 200~\degss (including the field discussed in this work) having deeper 220~GHz
data.
Following completion of the SPT-SZ survey, the 
receiver will be reconfigured for polarization sensitivity and will
image a subset of the SPT-SZ survey area to significantly lower noise levels.

While three-frequency data from the complete 2500~\degss survey
will eventually be released, the first public map release from the SPT-SZ
survey consists of data taken during the 2008 season, when the SPT receiver was primarily sensitive in the 150 and 220~GHz bands.  This first release presents maps covering approximately
95~\degs observed in those two bands. 
The area covered by these maps is referred to as the \fivehs field, named for the  
J2000 coordinates of the approximate field center.  
This was the first large field mapped to survey 
depth by the SPT and was centered on 
a 45~\degss subregion optically surveyed by the Blanco Cosmology 
Survey\footnote{http://cosmology.illinois.edu/BCS} (BCS, Desai et al., in prep.).
The BCS data were collected before 
the SPT was deployed, in anticipation of using the combination of 
optical and mm-wave data for joint galaxy cluster analyses.
The center of the BCS region and, hence, the center of the 
\fivehs field, is R.A. $82.7^\circ$, decl. $ -55.0^\circ$.
The \fivehs field has
been studied in detail to extract power spectrum measurements
\citep{lueker10, hall10, shirokoff11, keisler11} and to produce catalogs of emissive
sources \citep{vieira10} and SZ-selected galaxy
clusters \citep{staniszewski09, vanderlinde10,williamson11}.

This paper presents the 2008 SPT maps of the \fivehs field
at 150~GHz and 220~GHz, discussing in detail the aspects of the
instrument response and data processing that are relevant for
interpreting and using the maps.  We present maps with two variations 
in filtering and
map projection, one designed for cluster-finding and
one designed for the detection and characterization of emissive sources.  
It is not possible to
summarize all relevant features of 
the maps in a simple set of data products without some loss of
information, which limits the use of these maps for certain types of
analysis.  In particular, the data products presented here are not
sufficient to produce an extremely accurate noise model, nor to 
perform a cross-spectrum analysis for estimating the CMB power
spectrum, nor to perform jackknife analyses to test for contamination
\citep[e.g.,][]{shirokoff11,keisler11}.
All of these would require all maps of individual observations of the
\fivehs field, which is an order-of-magnitude larger set of data products
than what we are currently releasing.  Individual-observation maps 
may be included in future releases.

Regardless of the intended use of the maps, we emphasize the
importance of understanding how the SPT instrument response, data
processing, mapmaking, and noise properties affect the signals of
interest.  To that end, after describing the instrument and observing
strategy in Sections~\ref{sec:instrument} and
\ref{sec:observations}, we focus the majority of the paper
(Sections~\ref{sec:response} and \ref{sec:processing}) on a detailed
discussion of these properties of the data.  The maps themselves are
presented in Section~\ref{sec:maps}, which also presents cross-checks of analyses
using these maps compared to previously published SPT analyses.  
Section~\ref{sec:conclusions} describes the data
products available online.   We provide an example calculation using
these data products in Appendix \ref{app:examplecalc}.

\section{Instrument}
\label{sec:instrument}
The SPT is a 10-meter off-axis Gregorian telescope located at the
National Science Foundation's Amundsen-Scott South Pole Station, where atmospheric
conditions are among the best in the world for mm and sub-mm
observations \citep[e.g.,][]{radford11}.  The receiver
images the sky using an array of transition-edge-sensor (TES) bolometers
read out using frequency-multiplexed Superconducting QUantum
Interference Device (SQUID) amplifiers.  A description
of the instrument design and performance can be found in
\citet{carlstrom11}. 
Here, we summarize the aspects of the
instrument design and performance that are most relevant for
understanding the data products in this release. 

The detector array in the SPT-SZ receiver is made up of
six wedge-shaped sub-arrays, each of which has 140 detector pixels
configured to 
observe in one of the 95~GHz, 150~GHz, or 220~GHz observing bands.\footnote{Each detector wedge has 161 potential bolometer channels, of
  which 140 are read out.}   These
three observing bands have been selected to coincide with ``windows'' of
high atmospheric transmission, and to optimize discrimination of the
tSZ spectral signature.  The observing bands are defined on the
low-frequency end by a circular 
waveguide coupled to each detector, and on the high-frequency end by
low-pass metal-mesh filters mounted above each detector wedge \citep{Ade06}.
Detectors across a given wedge have similar band-pass profiles, with
an average bandwidth of 35~GHz for the 150~GHz band and 44~GHz for the
220~GHz band in the 2008 receiver. Details on the measured bands for
2008 are given in Section~\ref{sec:bands} and discussed further in
Bleem et al.~(in prep.).

Data from each individual bolometer channel are digitized at 1~kHz,  
then digitally low-pass filtered and down-sampled to 100~Hz before
being written to disk.  We refer to the resulting data stream as
time-ordered data (TOD).
The optical
time response of each detector is approximately described 
by a single-pole low-pass filter with time constants varying between 10 and
  30~ms for different detectors.  Measurements of the time response
  functions are presented in Section~\ref{sec:timeconstants}.  The measured
time response is deconvolved from the data during analysis, and additional
anti-aliasing filters are applied when the data are binned to create
maps, as described in Section~\ref{sec:filtering}. 

Each detector's beam, defined as the response of the detector as a function
of angle to a point source on the sky,
is determined by the combination of a conical
feedhorn above the detector and 
the optical design of the telescope \citep{padin08}.  The main lobes of the beams are
well described by two-dimensional Gaussians with
average full-widths at half maximum (FWHM) of 1.15 and 1.05 arcmin at
150~GHz and 220~GHz, respectively.  Individual detector beam profiles
vary, primarily depending on their placement in the focal plane. 
However, a single average beam for each frequency is appropriate for
characterizing the effective angular response in the final maps.
Beam measurements are presented in 
Section~\ref{sec:beams}.   

Noise in the SPT-SZ data comes from four main sources:  1) noise due
to statistical fluctuations in photon arrival time, 2) noise
intrinsic to the detectors, 3) noise from the readout system, and 4)
brightness temperature fluctuations in the atmosphere, mostly due to
inhomogeneous mixing of water vapor. 
The first two components are expected to contribute essentially
``white'' noise (i.e., equal amplitude at all temporal frequencies).  
Readout noise has a white noise component as well as
a ``$1/f$'' component, with power decreasing as temporal frequency
increases. The spatial power spectrum of atmospheric fluctuations 
increases steeply with increasing spatial scale \citep[e.g.,][]{bussmann05}, 
leading to noise in the TOD that rises steeply at low temporal frequencies.  
The atmosphere dominates the SPT noise at 
frequencies below 1~Hz, while the photon noise dominates at the higher
temporal frequencies that correspond to the signal region for cluster and point
source science.     
Typical single-detector noise-equivalent-temperatures (NETs) expressed in
CMB units 
range from approximately \netonftylow \ to 
\netonftyhigh~$\mu$K$\sqrt{\mbox{s}}$ at 150~GHz and 
\netttwntylow \ to \netttwntyhigh~$\mu$K$\sqrt{\mbox{s}}$ at 220~GHz for the data presented
here.  These NETs are estimated 
from the detector noise power spectra in a band corresponding to
multipole $\ell \sim 3000$ in the scan direction, using the absolute
temperature calibration described in Section~\ref{sec:abscal} of this
release.  Given
the typical numbers of detectors with good performance during 2008
observations, the total mapping speed was approximately
\netonftyarray~$\mu$K$\sqrt{\mbox{s}}$ at 150~GHz and 
\netttwntyarray~$\mu$K$\sqrt{\mbox{s}}$
at 220~GHz for the 2008 season.

\section{Observations} 
\label{sec:observations}

The primary observing mode for the SPT 
is scanning across the sky at constant elevation.  Because the SPT 
is located within $1$~km of the geographic South Pole, this corresponds
almost exactly to scans at constant declination.  Complete observations of 
a field are assembled from many consecutive scans at stepped
positions in elevation. 
Throughout each roughly 36-hour cryogenic cycle, we 
perform multiple short calibration measurements interleaved with
the field observations.  These calibration measurements include 
observations of a small chopped signal from a non-aperture-filling thermal source, 
two-degree scans in elevation, and observations 
of the Galactic HII regions RCW38 and MAT5a (NGC3576), which are common
calibration sources for mm-wave CMB experiments
\citep{puchalla02,coble03,kuo07}.  This set of regular
calibrations allows us to characterize instrument response and
monitor detector performance, as described in Section
\ref{sec:response}.

Between February 13 and June 5 2008, 421 observations were performed
of the  $\sim$95~\degss \fivehs field, each taking about two hours of
observing time.  
Each complete observation comprised 176 
constant-elevation scans across the field, with elevation offsets
of 0.125 degrees between pairs of scans back and forth across the field.
Twenty different initial starting elevations were used for successive
observations, at offsets of 0.005 degrees. 
Variations in the
starting position of successive observations enhance the uniformity of
coverage in combined maps.  
Approximately half of the observations of this field were performed
with an azimuth scanning speed of 0.44~deg/s, with the remaining observations
performed at 0.48~deg/s.

\section{Characterization of Instrument Response}
\label{sec:response}
\subsection{Observing bands}
\label{sec:bands}
The SPT spectral bandpasses are measured using a beam-filling Fourier Transform
Spectrometer (FTS), as described in Bleem et al.~(in
prep.).  Transmission spectra were measured for 
$\sim$50\% of the detectors on each of the six detector wedges. For a
given wedge, the detector transmission spectra are highly uniform, with
well-defined band edges at low and high frequencies that are set
by a precision-machined circular waveguide and a common metal-mesh low-pass
filter.  For detectors in the same wedge, the band center has an rms
variation of $\sim$1\%. 
For each band, we construct an average response by weighting each detector's
transmission spectrum by the inverse square of the detector's NET.
Uncertainties in the final spectra in each band are dominated by the absolute frequency
calibration of the FTS.  We can verify this by comparing the measured vs.~expected 
location of the low band edge (due to the circular waveguide cutoff), 
and we estimate this absolute frequency scale to be accurate to 0.3~GHz.
The 150 and 220~GHz transmission spectra, averaged over all detectors
in a given band, are available
for download (see Section~\ref{sec:conclusions}), expressed as the 
response to a beam-filling, flat-spectrum ($I(\nu)=\mathrm{constant}$) source,
with the peak transmission normalized to unity. 

In Table~\ref{tab:bands}, we give the band center and effective
bandwidth for the 
150 and 220~GHz bands.  We have defined the band center to be
\begin{equation}
\nu_\mathrm{cen} = \frac{\int \nu f(\nu) d\nu}{\int f(\nu) d\nu},
\end{equation}
where $f(\nu)$ is the transmission spectrum averaged over all 
detectors in a given observing band, and the
effective bandwidth 
is defined to be $\int f(\nu) d\nu$.  

To convert the SPT maps from CMB
temperature units to intensity, one must consider both the spectrum of the
source and the spectrum of the CMB.  For a source with a spectrum $I(\nu) = I_0 S(\nu)$, this
conversion factor is
\begin{equation}
\frac{I_0}{\Delta T} = \frac{\int A\Omega(\nu) \frac{dB}{dT}(\nu,T_{\rm CMB}) f(\nu) d\nu}{\int A\Omega(\nu) S(\nu) f(\nu) d\nu}, 
\end{equation}
where $A\Omega(\nu)$ is the telescope throughput, or etendue, and
$\frac{dB}{dT}(\nu,T_{\rm CMB})$ is the 
differential change in brightness of the CMB for a change in temperature.  For
beam-filling sources in a single-mode system such as the SPT-SZ receiver, 
$A\Omega(\nu) = c^2/\nu^2$.  

Spectra of astrophysical sources in mm-wave bands are typically approximated 
as power laws, such that $I(\nu) = I_0 (\nu/\nu_0)^{\alpha}$.  In Table 
\ref{tab:bands}, we give example conversion factors between CMB units and 
$\mathrm{MJy}/\mathrm{sr}$ for beam-filling sources with $\alpha$ values typical of 
some common mm-wave source families.  The conversion factors are quoted for 
$\nu_0$ equal to the nominal band center, i.e., either 150 or 220~GHz.  For 
comparison, we also quote the conversion factor that we would obtain if 
our bands were infinitely narrow and centered on the nominal band center.
(The conversion factor in this case is simply 
$\frac{dB}{dT}(\nu,T_{\rm CMB}) \times 10^{20}$, reflecting the definition of 
$1 \ \mathrm{MJy} = 10^{-20} \ \mathrm{W} \ \mathrm{m}^{-2} \ \mathrm{Hz}^{-1}$.)

To convert a measured temperature fluctuation in a CMB map to an
equivalent thermal SZ Comptonization or Compton-$y$ parameter 
\citep[e.g.,][]{carlstrom02}, one simply divides the measured $\Delta T$ 
by the mean CMB temperature and a frequency-dependent thermal 
SZ factor.  For delta-function bands, this factor is equal to

\begin{equation}
\fsz(\nu) = \left (x \frac{e^x+1}{e^x-1} - 4 \right ) (1 + \delta_\mathrm{\tiny{SZ}}(x,T_e)),
\end{equation}

where $x=h \nu / k_B T_{\rm CMB}$, $\delta_\mathrm{\tiny{SZ}}(x,T_e)$ is
a small relativistic correction \citep[e.g.,][]{nozawa00}, and $T_e$ is the electron 
temperature of the cluster.  For SPT bands and a beam-filling source, 
the effective band-averaged \fsz \ is equal to

\begin{equation}
\langle \fsz \rangle = \frac{\int \nu^{-2} \fsz(\nu) \frac{dB}{dT}(\nu,T_{\rm CMB}) f(\nu) d\nu}{\int  \nu^{-2} \frac{dB}{dT}(\nu,T_{\rm CMB}) f(\nu) d\nu}.
\end{equation}

In Table \ref{tab:bands}, we give values of effective, band-averaged \fsz \ for 
two values of $T_e$ (0 and 8~keV) for both the real bands and 
the delta-function approximations.

\begin{table*}[]
\begin{center}
\caption{SPT Average Band Properties}
\label{tab:bands}
\begin{tabular}{lcccc}
\hline
\multicolumn{1}{c}{Property} &
\multicolumn{2}{c}{150~GHz} &
\multicolumn{2}{c}{220~GHz} \\
\multicolumn{1}{c}{} &
\multicolumn{1}{c}{real} &
\multicolumn{1}{c}{delta-fn.~approx.} & 
\multicolumn{1}{c}{real} &
\multicolumn{1}{c}{delta-fn.~approx.} \\
\hline
Band center (GHz) & $   153.4$ & $150$ & $   219.8$ & $220$ \\
Bandwidth (GHz) & $    35.2$ & $-$ & $    43.7$ & $-$ \\
\hline
\multicolumn{1}{l}{Conversion factors for power-law spectra} & \\
~Radio, $\alpha=-0.5$ (MJy sr$^{-1}$ K$^{-1}$) & $   396.3$ & $   398.6$ & $   476.7$ & $   483.7$\\
~Rayleigh-Jeans, $\alpha=2$ (MJy sr$^{-1}$ K$^{-1}$) & $   389.6$ & $   398.6$ & $   487.7$ & $   483.7$\\
~Dusty, $\alpha=3.5$ (MJy sr$^{-1}$ K$^{-1}$) & $   375.6$ & $   398.6$ & $   487.5$ & $   483.7$\\
\hline
\multicolumn{1}{l}{Effective \fsz} & \\
~$T_e=0$ & $  -0.923$ & $  -0.954$ & $   0.006$ & $   0.038$\\
~$T_e=8$~keV & $  -0.876$ & $  -0.903$ & $  -0.046$ & $  -0.018$\\
\hline
\end{tabular}
\end{center}
\end{table*}

\subsection{Detector time constants}
\label{sec:timeconstants}

The temporal response function of SPT detectors over the signal band
of interest for this work ($\le 10 \ \mathrm{Hz}$) can be described by a single-pole low-pass
filter, with each detector characterized by a
single time constant.  The time constants are measured periodically
using the chopped thermal calibrator.  The time
constants are estimated by fitting the amplitude and phase response of
each detector using a sequence of chopper frequencies from 5 to 10~Hz. Time constants
are verified using fast scans of the detectors across bright
astrophysical sources.  Maps constructed using only left-going
scans can be subtracted from maps constructed with only
right-going scans to verify that residual time constant errors do not
contribute significant spurious signal in the combined maps.  Exactly 
these tests were performed in CMB power spectrum analyses of SPT 
data including the \fivehs field \citep{lueker10,shirokoff11,keisler11}, 
and no spurious signal was found.
The time constants do
not change significantly over a season or with receiver temperature over
the range of data used in the final maps,
so for analysis 
purposes a single time
constant parameter is associated with each detector for all
observations during a given season.   Time constants for the 2008
receiver configuration vary between about
10 and 30~ms across the detector array, with median
values of 19~ms and 17~ms for 150~GHz and 220~GHz detectors, respectively.

\subsection{Pointing Reconstruction and Astrometry}
\label{sec:pointing}

The real-time pointing model used to control the telescope is
initially calibrated using optical star cameras mounted on the
telescope structure, as described in more detail in \citet{carlstrom11}.  The
pointing reconstruction of the mm-wave data is then calculated offline using
daily measurements of Galactic HII regions and information from
thermal, linear displacement, and tilt sensors.  

During each cryogenic cycle, full observations are performed of the
HII regions RCW38 and MAT5a.  For each of these observations, the
response of each detector is fit to a scaled, translated version of
a template image of the HII region.  After any modification of the
focal plane or optical configuration (typically once a year), a set of
the RCW38 observations is used to measure each detector's pointing
offset relative to the telescope boresight.  The daily observations of
both RCW38 and MAT5a throughout the observing season are used to
constrain the pointing model over time, and the RCW38 measurements
contribute to estimation of relative calibrations, as described in
Section~\ref{sec:relcal}.
After the pointing model has been corrected using the HII region
and telescope sensor information, random errors 
in the pointing reconstruction of 
roughly $7\arcsec$ rms remain from observation to observation.
These random errors contribute to the
width of the effective beam in the final coadded maps, as described in the following section.

The absolute astrometry of the final coadded maps
was initially
calibrated by comparing the SPT positions of a handful of sources to
those sources' positions in the 
843~MHz Sydney University Molongolo Sky Survey 
catalog \citep{mauch03}.  This calibration was expected to be
accurate at the 10\arcsec \ level.
The recent publication of the 
the Australia Telescope
20~GHz Survey (AT20G) catalog \citep{murphy10} provides
an even more accurate astrometric calibration.  The astrometry in
the AT20G catalog is tied to VLBI calibrators and is accurate at the
1\arcsec \ level.  Using 17 sources in the SPT 150~GHz data, we find
very low scatter between SPT and AT20G positions.  We do, however, see
a small ($<10\arcsec$) but statistically significant mean offset from the
AT20G positions.  We correct this offset by simply changing the
definition of the field center for the \fivehs maps from its nominal
value of R.A. $82.70000^\circ$, decl. $ -55.00000^\circ$ to
R.A. $82.70247^\circ$, decl. $ -55.00076^\circ$.  After this
correction has been applied, the SPT positions of the 17 sources agree
with the AT20G positions to better than 1\arcsec \ in the mean, with
arcsecond-level scatter.  An estimated uncertainty of 2\arcsec \ in
each dimension 
accounts for statistical uncertainty in the SPT source detections and
systematic uncertainties due to potential offsets in the source
centers at the ATCA and SPT observing bands and potential offsets
between the SPT 150~GHz and 220~GHz maps.

\subsection{Beams}
\label{sec:beams}

A thorough understanding of the detector beams, or angular
response functions, 
is critical for interpreting the signals in SPT maps.
Sky signals are convolved with these functions in the
process of observation, leading to the extended appearance of point
sources in the maps and the suppression of small-scale power in the
angular power spectrum.  The structure of the SPT beams can be
characterized as a Gaussian main lobe 
out to a radius of 1 arcmin, near sidelobes at
radii between 1 and 5 arcmin, and a diffuse, low-level sidelobe at
radii between 5 and 40 arcmin relative to the beam center.

Dedicated observations of planets are valuable for measuring the far
sidelobes,
since planets are bright enough to adequately probe the tails of the
response function.  Planet observations are less useful for
studying the main beam, however, because of the limited dynamic range of the
detectors, and because the planets are extended sources.  We use
observations of the brightest quasar in each survey field to
characterize the main beam shape.  We reconstruct a two-dimensional
profile of the beam by stitching together measurements of the ``inner
beam'' within a 4-arcmin radius, and an ``outer beam'' covering
radii from 4 arcmin to 40 arcmin.  

The outer-beam measurements are based on 7 dedicated observations of Venus
performed in March of 2008, and 1 dedicated observation of Jupiter performed
in August 2008. The observations consisted of a sequence of azimuth scans with 0.5 
arcmin elevation steps between scans.
 Because the detectors are saturated when they observe
the planets directly (and require several detector time constants to recover),
only data from the first half of each scan is used.  For Jupiter
observations, the impact of the Jovian satellites, which is very small
to begin with, is mitigated by
subtracting a template based on their known locations.  The 
Venus and Jupiter data are filtered to remove atmospheric noise and CMB
fluctuations, with the
locations of the planets masked in the filtering. The average
scan-synchronous signal, as measured at distances larger than 40
arcmin from the planet, is subtracted from the maps.

The inner beam shapes for
this data release are
measured using a bright quasar that appears   
in the \fivehs field itself and a bright quasar that appears in the
other field observed by SPT in 2008, the \twentythreehs field.  We have
no evidence that the beam shape differs between these two fields.
A small map is constructed around the
brightest source in each field.  The source
is masked to a radius of 5 arcmin and
filtering is applied  to remove atmospheric noise and CMB.  The
residual CMB and noise in the central beam region leads to an
approximately constant offset to the absolute response in this map.  The inner
beam profile based on the quasar map is then stitched together with
the outer beam profile from Jupiter, with an iterative procedure that
uses the Venus maps to determine the scaling between the two
components and to correct for the constant offset in
the inner maps. 

This procedure leads to a composite two-dimensional beam profile that
describes both the main beam and far-sidelobe response of the
instrument.  In addition, because the inner-beam measurement is based
on the final coadded maps of the fields, it encapsulates
the contribution of several-arcsecond random pointing variations to the effective
angular response in the final maps.  Pointing variations
increase the effective beam width by approximately 3\% and 5\% at
150~GHz and 220~GHz, respectively.

\begin{figure*}[ht]

\begin{center}
\subfigure[150~GHz inner beam]{\label{fig:inner_beams_150} 
  \includegraphics[width=3in]{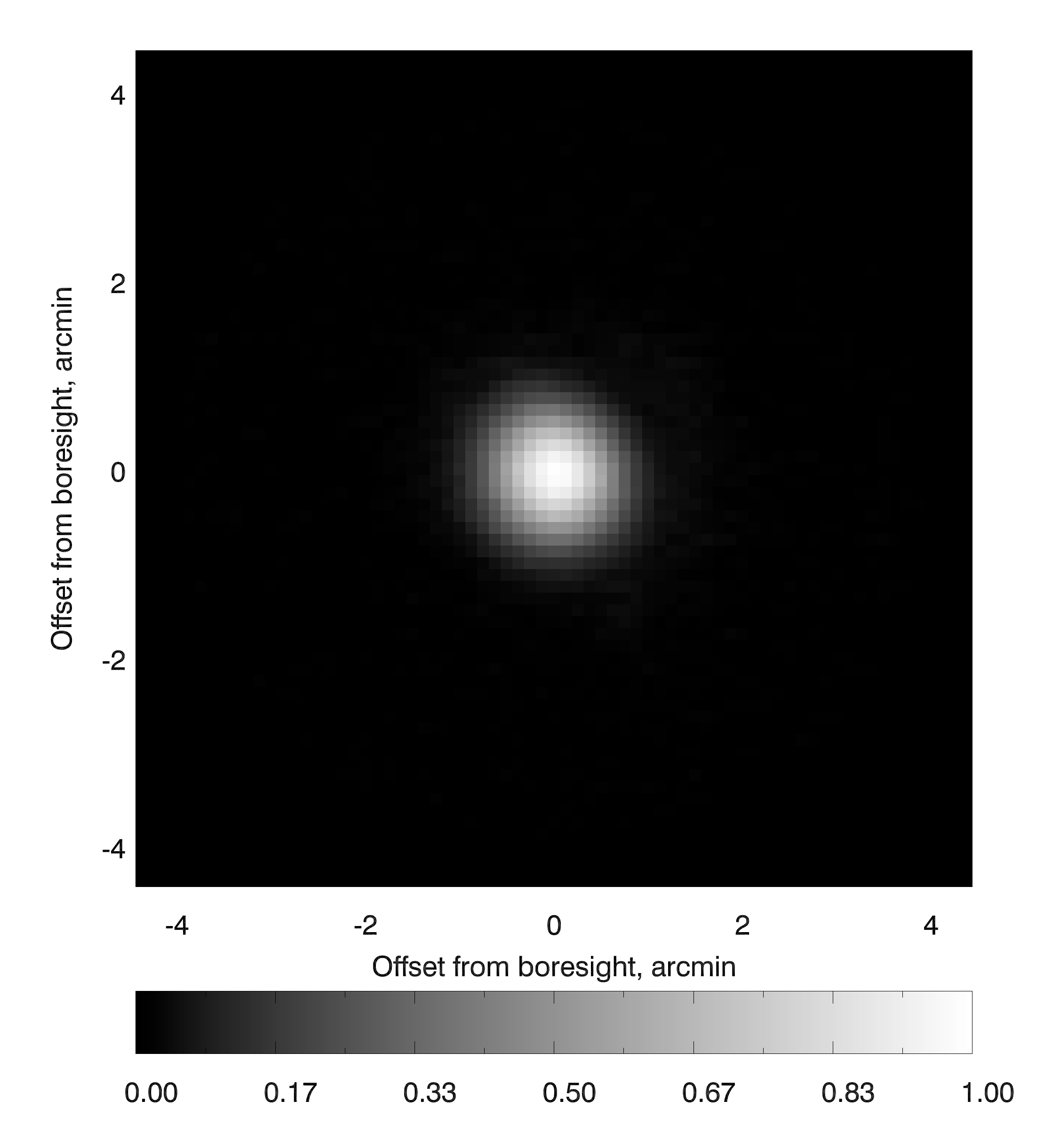}}
\subfigure[220~GHz inner beam]{\label{fig:inner_beams_220}
  \includegraphics[width=3in]{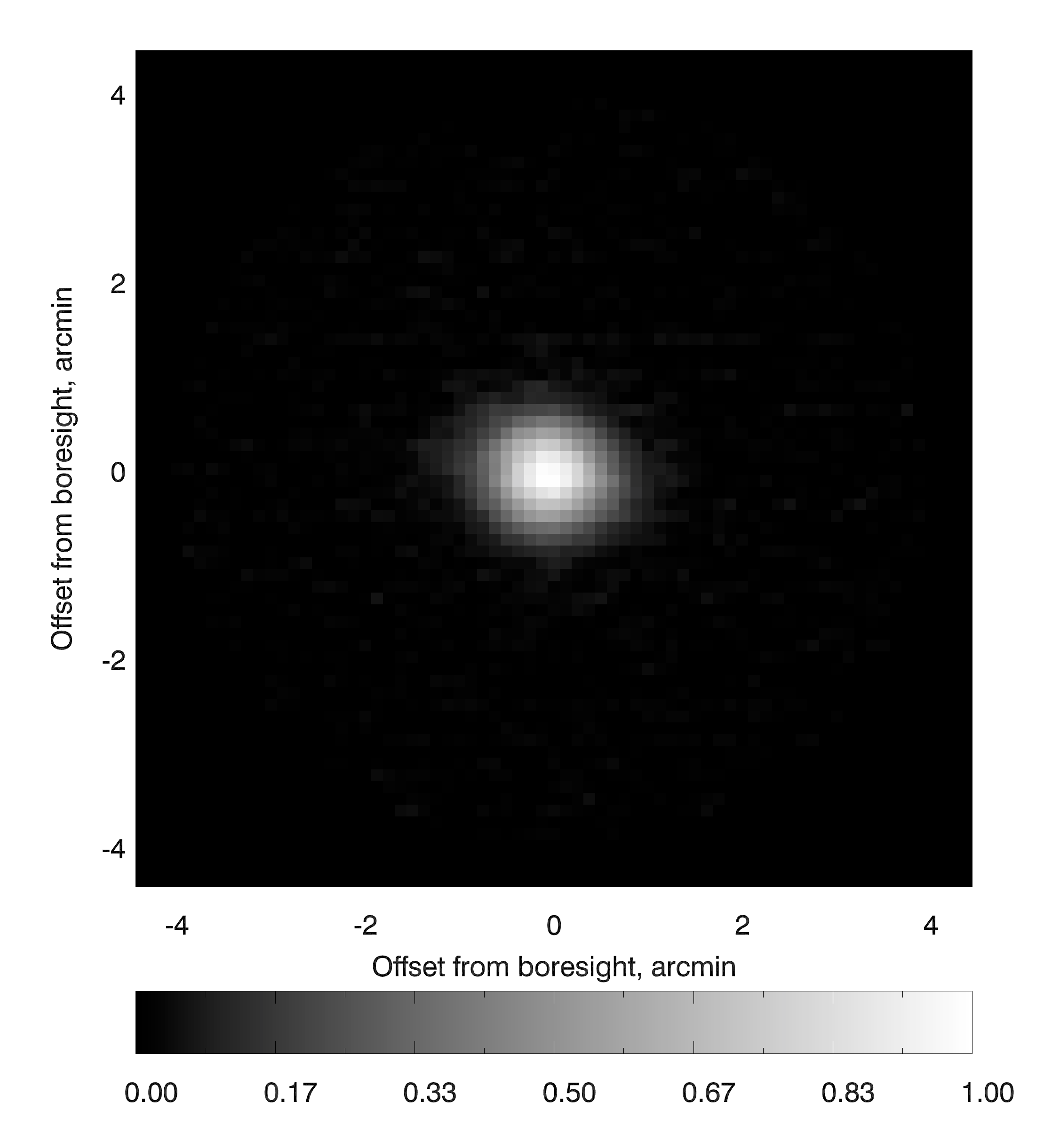}}
\subfigure[150~GHz outer beam]{\label{fig:outer_beams_150} 
  \includegraphics[width=3in]{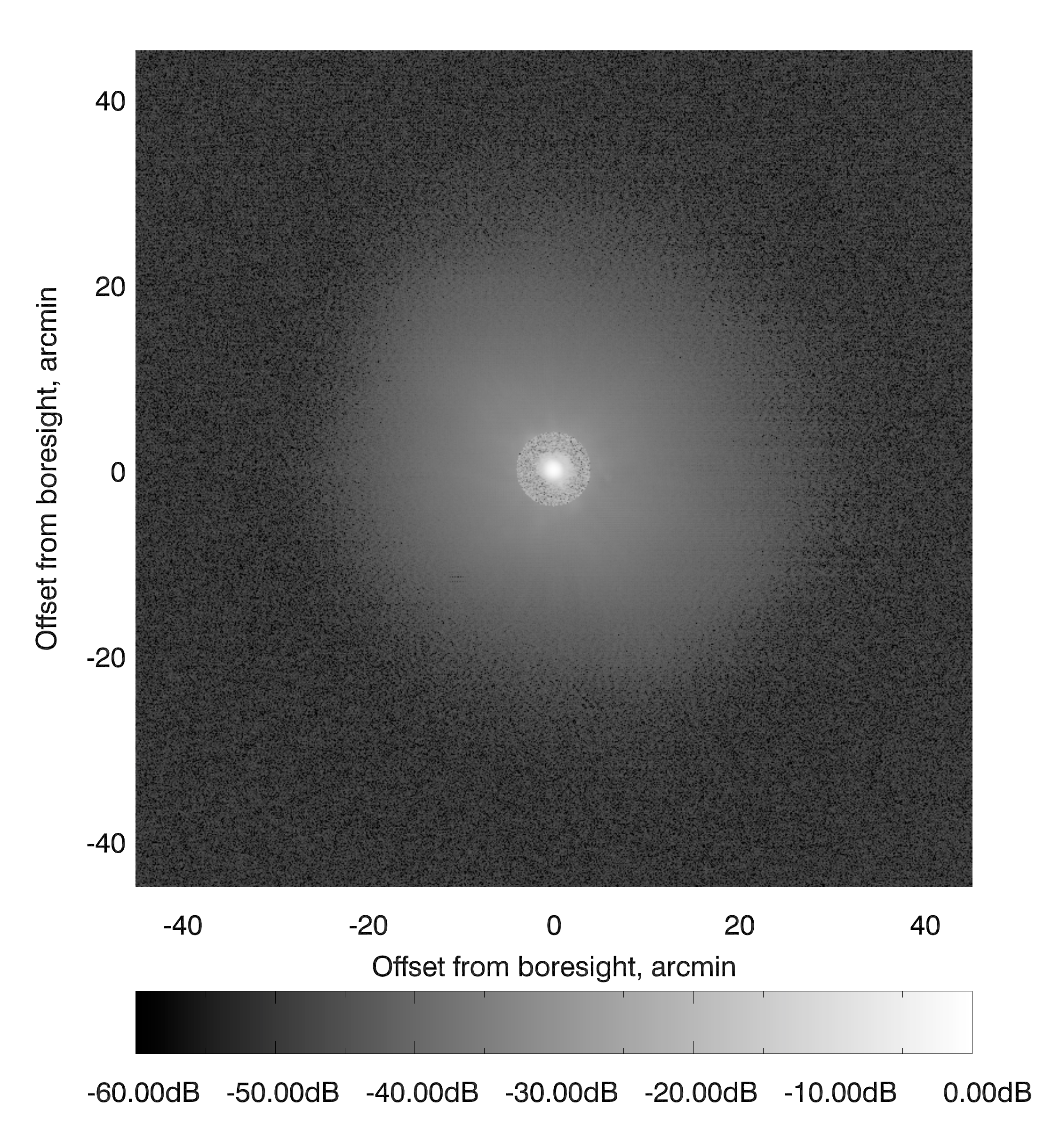}}
\subfigure[220~GHz outer beam]{\label{fig:outer_beams_220}
  \includegraphics[width=3in]{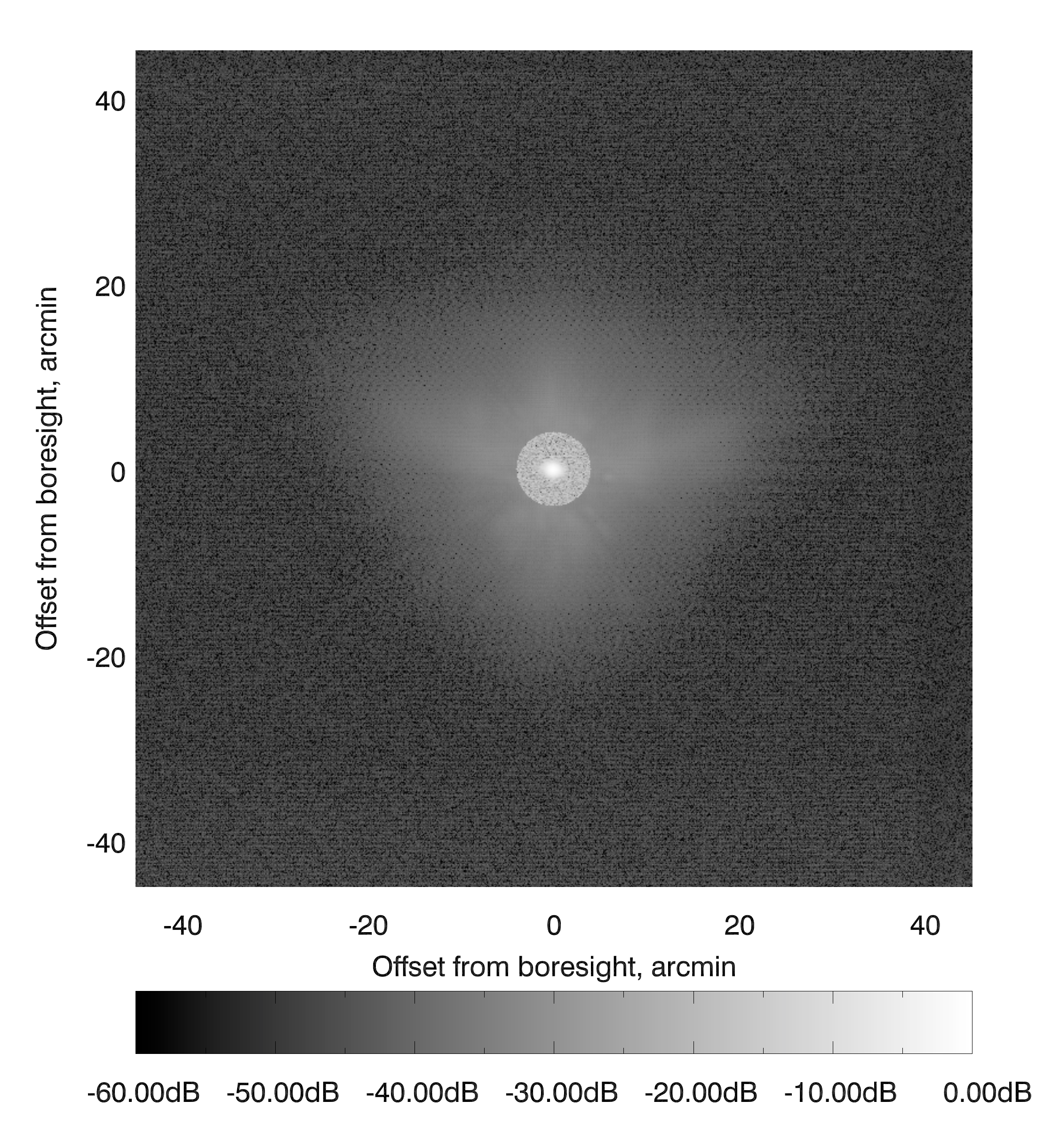}}

\end{center}
\caption{\label{fig:beams} Two dimensional average beam functions for
  the \fivehs field.  Panels (a) and (b) show the 150~GHz and 220~GHz
  two-dimensional beam profiles on a linear scale, emphasizing the
  structure of the ``inner beam.''  Panels (c) and (d) show the beam
  functions on a logarithmic scale out to much greater radii,
  emphasizing the structure of the ``outer beam,'' and showing the
  stitching of the two beam estimates at a radius of 4 arcmin.
  Note that in panels (c) and (d) the absolute value has been taken in
  order to display the image on a logarithmic scale, but this visually
  exaggerates the appearance of the noise in the innermost 4
  arcmin of the beam.}
\end{figure*}

The inner and
outer beam profiles for both bands are presented in Figure
\ref{fig:beams}.  The composite beam
maps are available for download, as described in Section~\ref{sec:conclusions}.

\begin{figure*}[ht]
\begin{center}
  \includegraphics[scale=0.68]{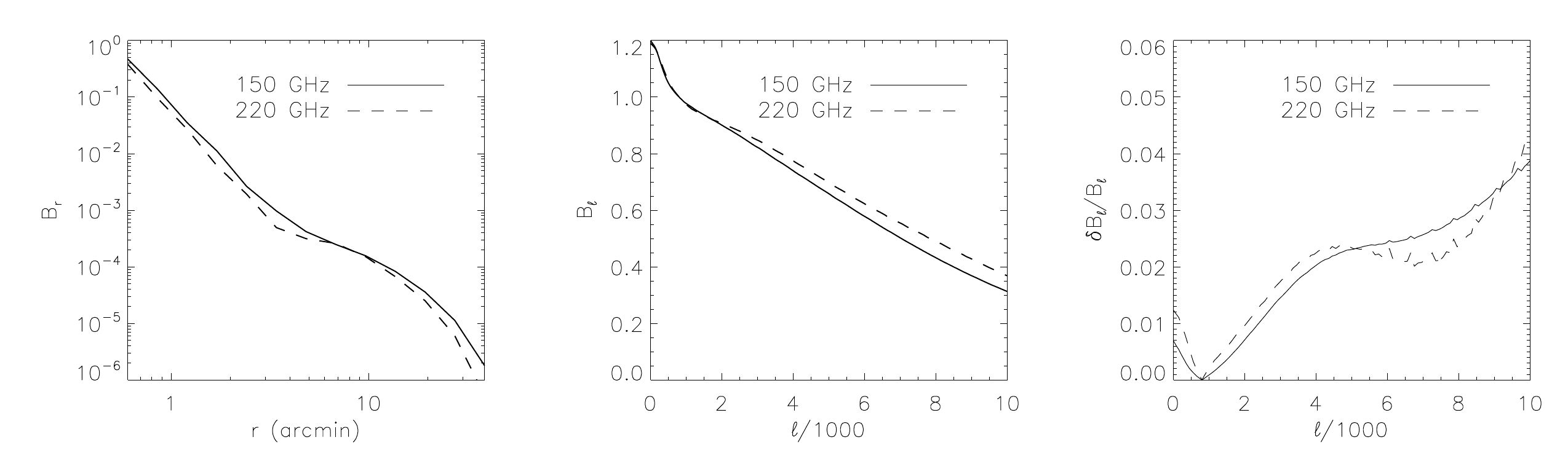}

\end{center}
\caption{Azimuthally averaged beam profiles (left), Fourier-space beam
  functions (center), and fractional uncertainties (right).  \label{fig:beam_profiles} }
\end{figure*}

In many applications, it is preferable to use a simplified
approximation to the beam profiles rather than the full
two-dimensional angular response functions. Azimuthally averaged beam
functions in map space and Fourier space are presented in Figure
\ref{fig:beam_profiles}. Using the flat-sky approximation, we
calculate the Fourier 
transform (FT) of the composite beam map, $B(\ell,\phi_\ell)$. From this,
we compute the azimuthally averaged beam function, 
\begin{equation}
B_\ell = \sqrt{\frac{1}{2\pi}\int |B(\ell,\phi_\ell)|^{2} d\phi_\ell}.
\end{equation}
There is a small bias ($<0.5\%$ fractional error at $\ell<10000$) in this estimate of
$B_\ell$ due to residual map noise, and we remove this bias. 

The normalization of $B(\ell)$ is somewhat arbitrary, 
in that it is degenerate with the absolute,
CMB-power-spectrum-based calibration factor described in Section
\ref{sec:abscal}.  Our CMB-power-spectrum-based calibration uses the multipole  
range $650 \le \ell \le1000$, so we choose to normalize $B(\ell)$ to 1 at $\ell=800$ to
minimize the correlation between beam uncertainty and calibration
uncertainty.

The uncertainty in our estimate of $B(\ell)$ arises from several statistical and systematic
effects, including residual atmospheric noise in the maps of Venus and
Jupiter, and the weak dependence of $B_\ell$ on the choice of radius
used to stitch together the inner and outer beam maps.  We consider
seven sources of uncertainty in total.  In
Figure~\ref{fig:beam_profiles} we show the quadrature sum of these
error estimates, which approximates the total uncertainty. The beam
functions are uncertain at the few percent level, and this uncertainty
increases mildly with increasing multipole number.

\subsection{Calibration}
\label{sec:calibration}

Averaging measurements from many
detectors within a given observation requires correcting for their
relative gains.  To average multiple single-observation maps together,
we also need to correct for inter-observation detector gain variations and
changes in atmospheric opacity that affect the calibration of the
whole array.  Section~\ref{sec:relcal} describes our procedure for
estimating detector-to-detector and day-to-day relative calibrations.
When all of these relative calibration factors are
applied, a single absolute calibration number relates the amplitudes
in the final map to physical units. Section~\ref{sec:abscal} describes
our procedure for estimating the final absolute calibration for each
map.  The physical units chosen for our maps are
K-CMB, expressing deviations from the average measured intensity as
equivalent temperature fluctuations in the CMB. 

\subsubsection{Relative Gain Calibrations}
\label{sec:relcal}

Two types of calibration observation are combined to account for
detector-to-detector relative gains, temporal gain variations, and atmospheric
opacity variations: 1) daily observations of the HII region RCW38,
and 2) observations of the chopped thermal calibration source that
illuminates the focal plane from the center of the telescope secondary mirror.
The thermal calibration source illuminates the focal plane through a 
small hole in the secondary mirror; hence, the effective source shape 
seen by the detectors is very different from that of sources on the sky, 
and varies over the focal plane.  

The thermal calibration source is observed many times per day,
and we use these observations to correct for variations in response 
to sky signal across the array (flat-fielding) and to correct for any gain drifts.
The first step in the relative calibration pipeline is to relate each detector's
response to the calibration source to its response to an astrophysical
source.  To do this, we assign an effective temperature to the calibration source 
for each detector, based on the season average of ratio between the 
response to the calibration source and the response to RCW38.  
The relative calibration for each detector over a single \fivehs 
observation is then based on the response to the calibration source
observation nearest that field observation.  To account for any day-to-day
drifts in the calibration source filament temperature or illumination
pattern, we correct the single-observation relative calibration number
for each detector using the wedge-averaged difference between that 
observation's calibrator response and the season average.  Similarly, 
to account for any changes in atmospheric opacity, we correct each 
detector's calibration using the wedge-averaged difference between 
response to the nearest RCW38 observation response and the 
season average.  The relative calibration factors applied to a given
detector are typically stable to within $2\%$ over the season.

\subsubsection{Absolute Calibration}
\label{sec:abscal}

The HII regions RCW38 and MAT5a can be used as absolute calibrators,
but their irregular shapes and the 
uncertainties in their absolute fluxes at mm-wavelengths limit the
precision of the calibration.  An RCW38-based absolute calibration has
been used for some previous work \citep{staniszewski09, williamson11,
 story11}, but here we adopt a refined absolute 
calibration based on 
comparisons of CMB angular power spectra produced from
SPT and WMAP data.  
The two
approaches give calibrations that are consistent within their uncertainties.

To calibrate the coadded maps to absolute temperature units, we
estimate the angular power spectrum of the CMB at 150~GHz using
the technique described in \citet{keisler11}.  The maps used in this 
estimation have been preliminarily calibrated using the RCW38-based calibration.
The resulting spectrum is then compared to the well-calibrated WMAP7 
power spectrum \citep{larson11} over the multipole range $650 \le \ell \le 1000$, 
and the final calibration to apply to the 150~GHz SPT maps is determined by 
requiring the inverse-variance-weighted ratio of the SPT and WMAP7 power
spectra to be unity over this range.
To reduce the contribution of SPT noise and
sample variance to the calibration uncertainty, we use all 2008 data
to derive the calibration for the \fivehs field.  We have looked for evidence 
of calibration differences between the 2008 data from the \fivehs field and
the rest of that season's data by checking the regular
observations of MAT5a, and we constrain this difference to be smaller 
than 1$\%$ in temperature.
A cross-power spectrum is constructed using the 150~GHz and 220~GHz maps, 
correcting for differences in beams and filtering between the two bands. 
 The ratio of this cross-power spectrum to the power spectrum at 150~GHz for 
multipoles dominated by primordial CMB signal
gives an estimate of the relative calibration offset between the two bands, 
which is used to transfer the 150~GHz absolute calibration to 220~GHz.

We estimate that the uncertainty in our 150~GHz absolute temperature calibration is
3.1\% using this calibration method.  The uncertainty in the 220~GHz
absolute calibration is estimated to be 6.9\%.  Note that since the
220~GHz calibration derives from the 150~GHz calibration, the two
uncertainties are correlated with a correlation coefficient of
approximately 0.5.

\section{Data Processing and Map-Making}
\label{sec:processing}
Raw time-ordered data are processed into maps in two
stages.  In the first (``pre-processing'') stage, the raw data from a
single observation of the field are calibrated, data selection cuts are
applied, and initial filtering and instrument characterization are
performed.  The same pre-processing pipeline is used for all SPT
survey data, regardless of the intended final use of the maps,
although there have been some minor modifications in the data
selection criteria and pre-processing algorithms over the history of
SPT data analysis.

In the second (``map-making'') stage, additional filtering is
performed on the pre-processed TOD and the data are binned into
single-observation maps used for final coadds. The filtering
applied in the map-making stage is typically customized for different
analyses, as is the choice of map projection.  As described in detail
below, we present two specific combinations of filtering and map
projection in this release.

We characterize the effect of the filtering and projection choices by
describing the Fourier-space properties of the maps.  We refer to
features in the Fourier plane using the corresponding angular
wavenumbers in radians {\bf k}.  The maps are oriented such that at
the center of the map, the $x$-direction corresponds to R.A. and the
$y$-direction corresponds to declination.  

\subsection{Data Selection and Pre-Processing}
\label{sec:preprocessing}

The first steps in the pre-processing pipeline are to characterize
detector performance, assemble measured response characteristics,
and calculate calibration factors and weights.  

Summary statistics are first calculated to characterize the receiver
setup and the TOD for each detector. Each detector's TOD for an entire
2-hour observation is used to estimate
the power spectral density (PSD) for that detector.  A fit to the noise PSD is then used to identify
line-like features that deviate from a
$1/f$ plus white noise profile.   Additional noise statistics are
calculated for the SQUIDs.

Next, a set of responsivity statistics is assembled
for every detector.  These include the amplitude of the detector's
response to the chopped thermal calibrator, the amplitude of its
response during a calibration observation of RCW38, and the amplitude of
its response to a 2-degree elevation scan performed before each field
observation.  These response parameters are used to calculate daily
gain calibrations for each detector, as previously described
in Section~\ref{sec:calibration}. They are also used to define data
selection cuts.

Preliminary cuts verify that each bolometer's 
bias voltage and readout configuration settings are
within the nominal ranges and that the TOD values for each channel remain
within the dynamic range of the digitizer.  Once these initial
cuts have been applied, bad bolometers are rejected with progressively
more and more stringent assessments of responsivity and noise.  

The first such cuts enforce a
minimum signal-to-noise response to the 
chopped calibration source, and a minimum response to the routine elevation scans.
Bolometers are also rejected if their response to the elevation scans
does not fit the expected modulation of the atmosphere. If a
detector's response to the chopped calibration source or the short elevation
scans is more than three standard deviations away from the median
response for detectors of the same band, it is also rejected.
Detectors are rejected if their PSDs exhibit wide line-like features
or too many lines.  After all of these cuts are applied, a final
pair of cuts rejects bolometers with calibration constants or noise
weights that deviate by more than a factor of three from the median for
each band.   The median numbers of detectors that pass cuts for the \fivehs field
observations (quoted in Section~\ref{sec:coadds}) reflect the bolometers
selected through this process.  

The bolometer data are parsed into scans, defined as temporally
contiguous periods of
constant-velocity azimuth scanning in either direction across the
field.  Cutting periods of time when the telescope is accelerating 
(at the ends of each scan) removes about 5\% of the total TOD for an observation. 
All data for a given scan are cut if the receiver
temperature exceeded the nominal operating value at any time during
the scan, if the telescope following error (absolute value of
the commanded position minus the position recorded by the encoders) 
exceeded 20\arcsec \ at any time during the scan, or if there
were data acquisition problems leading to bolometer data or pointing
data drop-outs during the scan.  Typically about 5\% of scans are cut
for these reasons.  Data for an individual bolometer in an individual
scan are also rejected if that bolometer shows evidence of step-function
features in its TOD (which can sometimes occur due to ``flux jumps'' in
the SQUIDs), if the SQUID or bolometer noise for that scan is
excessive, or if any of the digitized SQUID data
approach the limits of 
the dynamic range.  A spike-finding algorithm identifies cosmic-ray-like
events in the TODs for individual detectors. If there are fewer
than five in a given scan, and the spike features are relatively small, we
remove them and interpolate over the gaps; otherwise, we flag and
ignore affected TODs for the duration of the scan. Typically, around 5\% of
otherwise well-performing bolometers are cut from each scan for one of these reasons.

Because a varying subset of detectors will sometimes show sensitivity
to the receiver's pulse-tube cooler (a phenomenon also seen in other
TES bolometer systems with pulse-tube coolers, e.g., \citealt{dicker09}), we apply a notch
filter to remove a small amount of bandwidth from all data during the
pre-processing stage.  Combining all data that pass selection cuts
for a given observation, we identify the fundamental frequency of the
pulse-tube cooler and notch-filter a conservative 0.007~Hz of
bandwidth around this frequency as well as any strong 
harmonics.  For the
2008 \fivehs field observations, the pulse tube operating frequency was 1.62~Hz.
If all harmonics of this frequency were always cut, this would
represent an absolute maximum of 0.4\% of the 
total bandwidth; the actual bandwidth cut is smaller since typically
only the first few harmonics are cut.  We have verified that this
notch filter has a negligible effect on further analysis, and we 
neglect it in further analysis steps.

After all pre-processing, the offline pointing model is used to calculate corrected
pointing positions to associate with every time sample, and the
pre-processed data are written into an intermediate data format for
further analysis.

\subsection{Filtering}
\label{sec:filtering}
After the pre-processing stage of data analysis, the TOD associated
with a given field observation is further filtered on a scan-by-scan
basis.   The processing of each scan 1) deconvolves the
detector temporal response functions, 2) removes high-frequency
temporal noise that translates to spatial scales smaller than the pixel
resolution in the maps, and 3) removes atmospheric noise while
minimizing the filtering of signal power. 

The time-constant deconvolution and low-pass anti-aliasing filters are
applied in a single Fourier-domain operation on each scan for each
detector.  The time-constant deconvolution takes
into account the measured time constants for every individual detector
(see Section~\ref{sec:timeconstants}).  A cut-off frequency of 25~Hz is used
for the low-pass filter applied to the TOD, chosen to limit noise on spatial scales
smaller than the 0.25-arcmin map pixels without suppressing power
on the spatial scales of the SPT beams.  

Prior to averaging the data into maps, additional time-domain
filtering is applied to remove atmospheric noise.  The atmospheric
contamination resides primarily at low temporal frequencies, so a
high-pass filter is an effective way to remove the
most contaminated data.  However, the application of a Fourier-domain
high-pass filter will also distort the appearance of bright point
sources in the final maps, introducing ``ringing'' patterns.  This
effect can be avoided by filtering in the time domain via 
fitting to slowly varying template functions 
and masking the locations of known bright point sources during the fit.
Point-source masking of this sort
has been performed for one of the sets of maps in this release, but not for
the other, as explained in Section~\ref{sec:thisfiltering}.  

In all variations of our analysis, we fit out a mean and slope as well as 
higher-order polynomials from each scan.  In some past work we have
subtracted Legendre polynomials and in others a series of 
sines and cosines (Fourier modes).  In either case the result is a
time-domain high-pass filter that allows for point-source masking as
needed.  For the maps in this release, we have fit Fourier modes up to 
a temporal frequency that corresponds to a 
spatial high-pass cutoff of $k = 300$ in the scan direction, or $k_x=300$ 
(because the scan direction
corresponds to the $x$ direction in our maps).

Atmospheric noise is highly correlated across the detector array.  By
taking this correlation into account, we can filter additional
atmospheric noise without affecting small-scale sky signal.  For
each TOD sample, we subtract the mean value across every detector in a
given wedge.  This filter can also be performed with the locations of 
bright point sources masked, and this masking is performed in one set of 
maps in this release (as with the masking in the time-series filters).

\subsection{Filtering Variations for Maps in this Release}
\label{sec:thisfiltering}

This paper presents two sets of maps, one tailored to cluster
analysis and one tailored to point-source analysis.\footnote{The 
majority of extragalactic emissive sources appear point-like in the
arcminute-resolution SPT maps, and we use ``point source" and 
``emissive source" interchangeably in this work.}
For cluster
analysis, the point sources in the maps are viewed as a foreground.
The locations of known point sources are masked in the
filtering for these maps in order to prevent artifacts from
ringing, which could affect cluster extraction.  The brightest 201
sources are masked in the filtering, with the top 9 masked to a radius
of 5 arcmin and the rest masked to 2 arcmin. The source list
used for constructing the mask was created by merging together all
sources detected at greater than $5\sigma$ significance in either
150~GHz or 220~GHz maps, from the published point-source catalog for
this field \citep{vieira10}, and adding any sources that lie outside the
field boundaries used for detection in that work but that are strongly 
detected in an analysis using all map pixels.

When the point sources themselves are the object of study, it is
preferable to filter them in the same way as the other parts of the
map, so that the effect of filtering on signal can be completely
characterized.  For the set of maps tailored to point source analysis,
the brightest point sources are not masked in the filtering.  We also
choose different map projections for the two sets of maps, as
described below.  All
other parameters of the data processing and filtering are equivalent
in both sets of maps.

\subsection{Map Projections}
\label{sec:projections}

After all data selection and filtering has been performed, we use the corrected pointing
information to average and bin the data into pixels in a
two-dimensional map.  For small areas of the sky, analysis of the maps
is greatly 
simplified by using a flat-sky approximation.  This
allows two-dimensional FTs to be used (rather than
spherical harmonic transforms) for analyzing signal 
power, noise, and instrument response as a function of angular scale.
However, any choice of map projection will lead to some
distortions of information in the map.  The character of these
distortions varies depending on the selected projection, meaning that
some analyses may be easier to perform on maps with one projection
while others may be easier with another.

In past work we have employed several different projection
schemes, of which two have been selected for presentation here. Past
SPT cluster-finding analysis has used the
Sanson-Flamsteed projection \citep[e.g.,][]{calabretta02}, which projects 
constant-elevation scans into pixel rows in the final map. This
projection choice
simplifies the characterization and treatment of signal filtering,
because the TOD processing effectively operates on map rows.  The
disadvantage of this projection is that it is not distance-preserving,
and sky features near the corners of the maps will therefore have
slightly distorted shapes (for example, a 1-arcmin circle will appear to 
have an ellipticity of $\epsilon \sim 0.12$ near the corner of the 
\fivehs map).  We have selected this projection for the
maps tailored to cluster-finding.

 The second projection employed here, for the maps tailored to
 point-source analysis, is the oblique Lambert equal-area
azimuthal projection.  For the typical size, shape, and center location of SPT maps, 
this sky projection preserves both distances and areas to high
accuracy \citep{snyder87}, which means that the effective beam shape will not vary
significantly for different locations in the map.  This is
particularly advantageous for reconstructing point source amplitudes
using variants of the CLEAN algorithm \citep{hogbom74}.  
The disadvantage
of this projection is that the effective filtering has a strong
position dependence, since the angle between the scan direction and
 the map pixel rows changes across the map. 
 For constant-declination scans, this angle, $\alpha$ can be expressed as:
 \begin{equation}
 \alpha = \tan ^{-1} (A/B),
 \label{eqn:rot}
 \end{equation}
 where
\begin{eqnarray*}
A&=&\gamma \sin \theta_0 \sin(\phi_0 - \phi)\times\\
&&  \left( \sin \theta_0 \cos \theta   + \sin \theta \cos \theta_0 \cos(\phi_0-\phi) \right) \\
&& + \cos \theta_0  \sin(\phi_0 - \phi),
\end{eqnarray*}	 
\begin{equation*}
		 B=\gamma \sin \theta_0 \sin^2(\phi_0-\phi) \sin \theta + \cos(\phi_0-\phi),
\end{equation*} 
and
\begin{equation*}
\gamma = \frac{0.5}{1+\cos \theta_0 \cos \theta + \sin \theta_0 \sin \theta \cos(\phi_0-\phi)}. 
 \end{equation*}
Here $\phi$ equals the R.A. in radians, while 
$\theta = \pi/2 - \mathrm{declination}$,
also in
radians. Unsubscripted variables represent the pixel location and the
subscript ``0'' denotes the field center.  

In Sections~\ref{sec:tfs} and \ref{sec:psds} we will discuss the impact of the choice of
projection on the signal and noise properties in these maps.  

\begin{figure*}[ht]
\begin{center}
\includegraphics[scale=0.95]{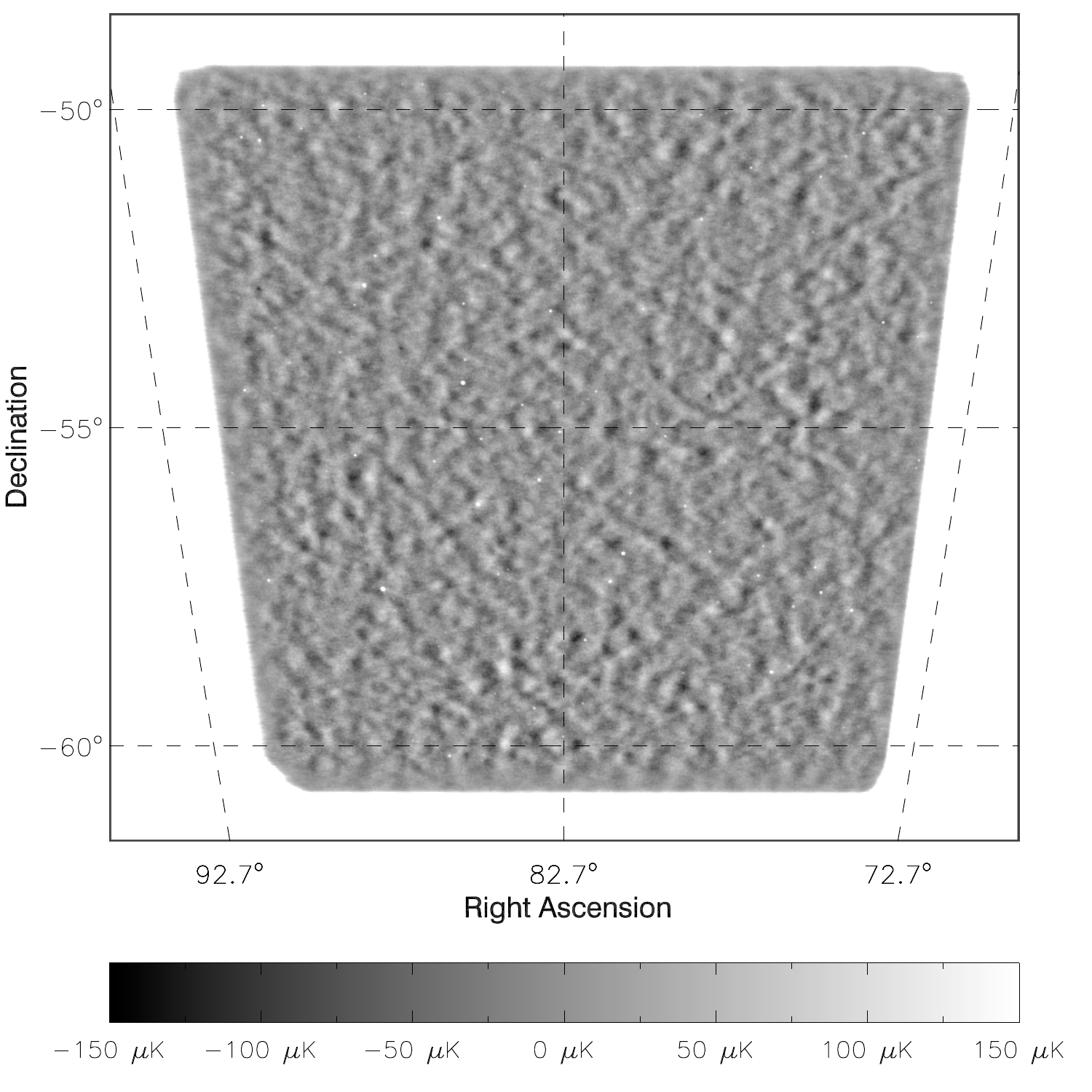}

\end{center}
\caption{Map of the \fivehs field at 150~GHz, using point-source
  masking in the data filtering and employing the Sanson-Flamsteed
  projection.  This is the filter and projection scheme tailored for performing cluster analysis.
  For display purposes, the map has been smoothed with a 1-arcmin FWHM Gaussian, 
  and high-noise regions near the boundary of the map have been masked.
{\label{fig:proj0_150}}}
\end{figure*}

\begin{figure*}[ht]
\begin{center}
\includegraphics[scale=0.95]{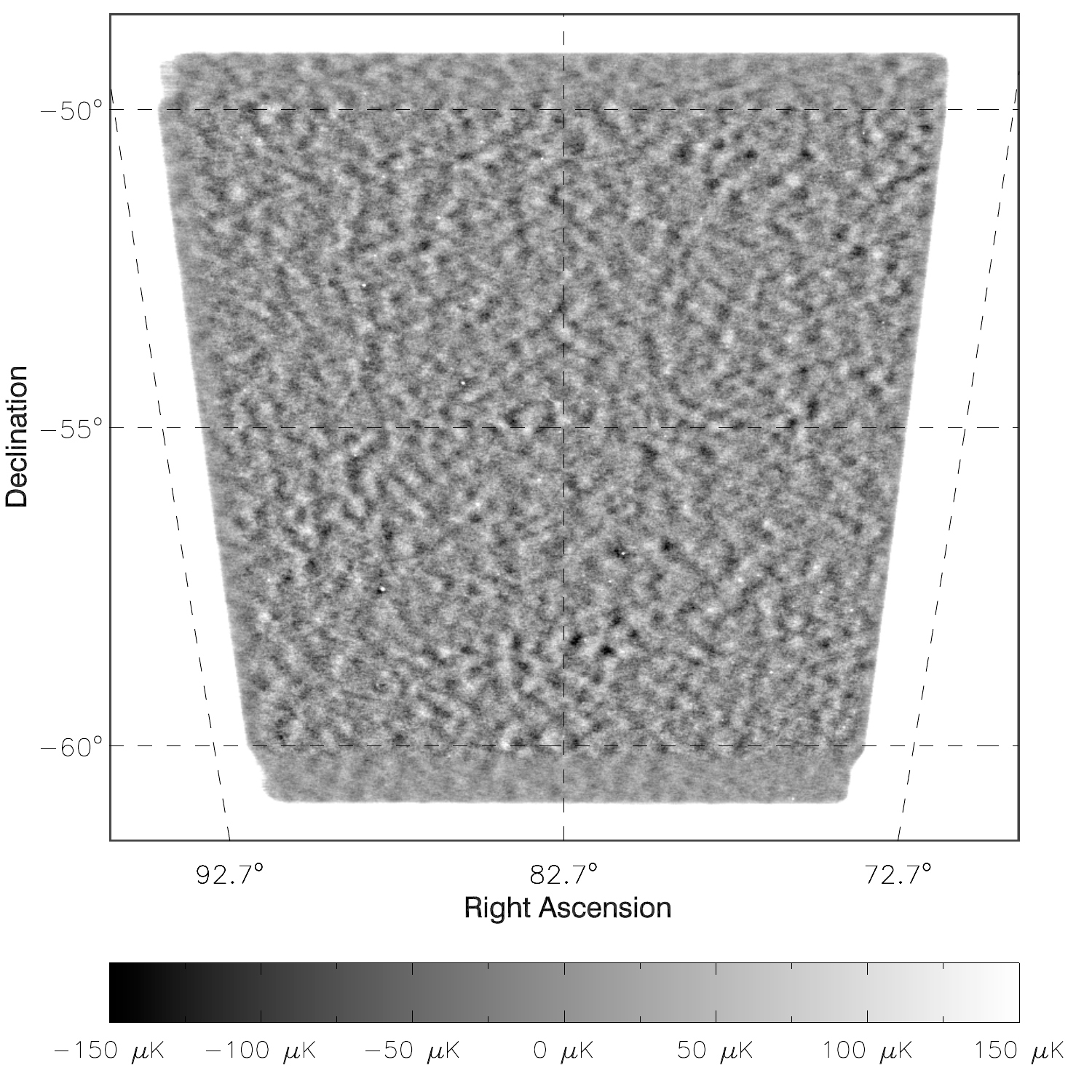}

\end{center}
\caption{Map of the \fivehs field at 220~GHz, using point-source
  masking in the data filtering and employing the Sanson-Flamsteed projection.  This is the filter and projection scheme tailored for performing cluster analysis.  
  For display purposes, the map has been smoothed with a 1-arcmin FWHM Gaussian, 
  and high-noise regions near the boundary of the map have been masked.
{\label{fig:proj0_220}}}
\end{figure*}

\begin{figure*}[ht]
\begin{center}
\includegraphics[scale=0.95]{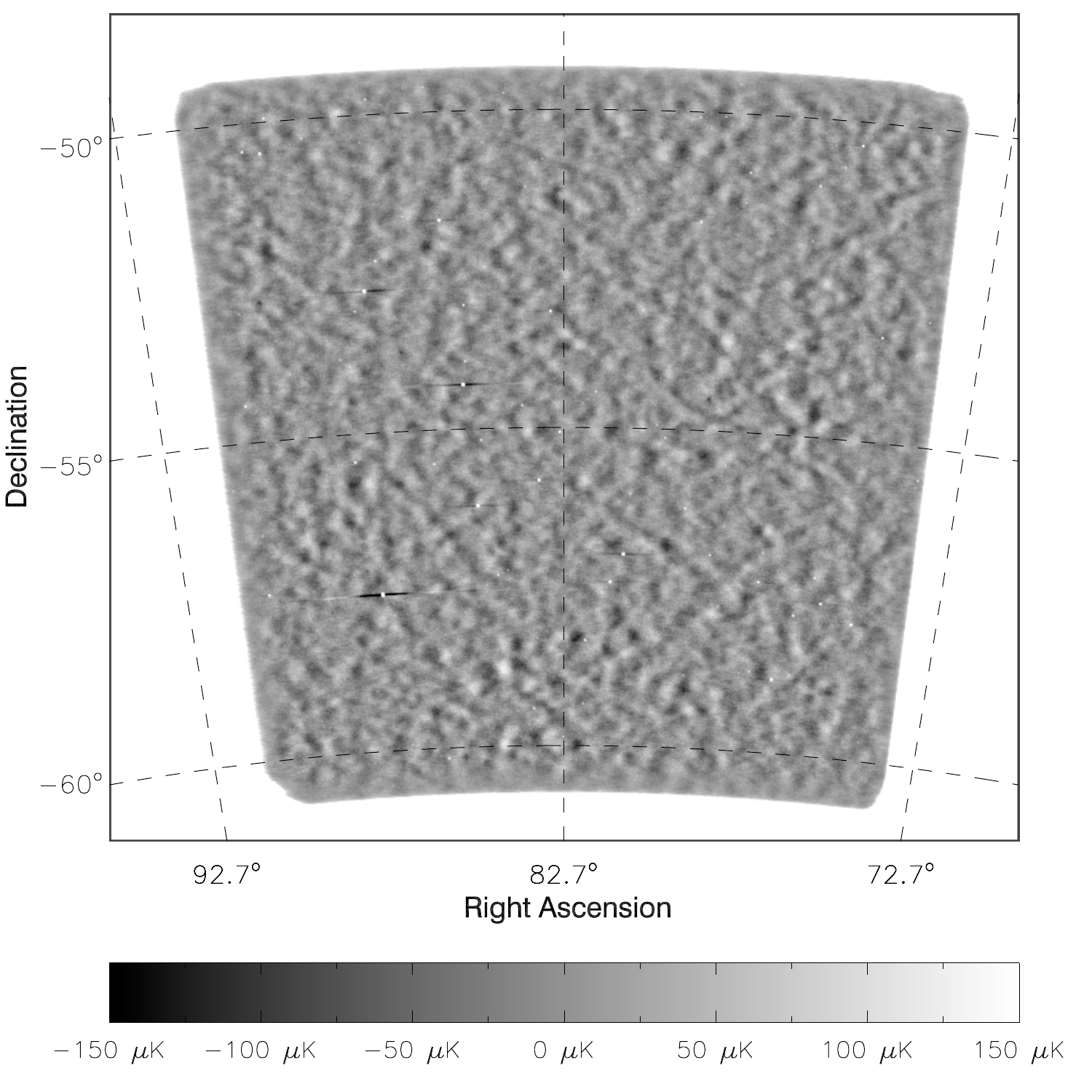}

\end{center}
\caption{Map of the \fivehs field at 150~GHz, with point sources
  unmasked during data filtering, using the oblique Lambert equal-area
  azimuthal projection.  This is the filter and projection scheme tailored for performing point-source analysis.  
  For display purposes, the map has been smoothed with a 1-arcmin FWHM Gaussian, 
  and high-noise regions near the boundary of the map have been masked.
{\label{fig:proj5_150}}}
\end{figure*}

\begin{figure*}[ht]
\begin{center}
\includegraphics[scale=0.95]{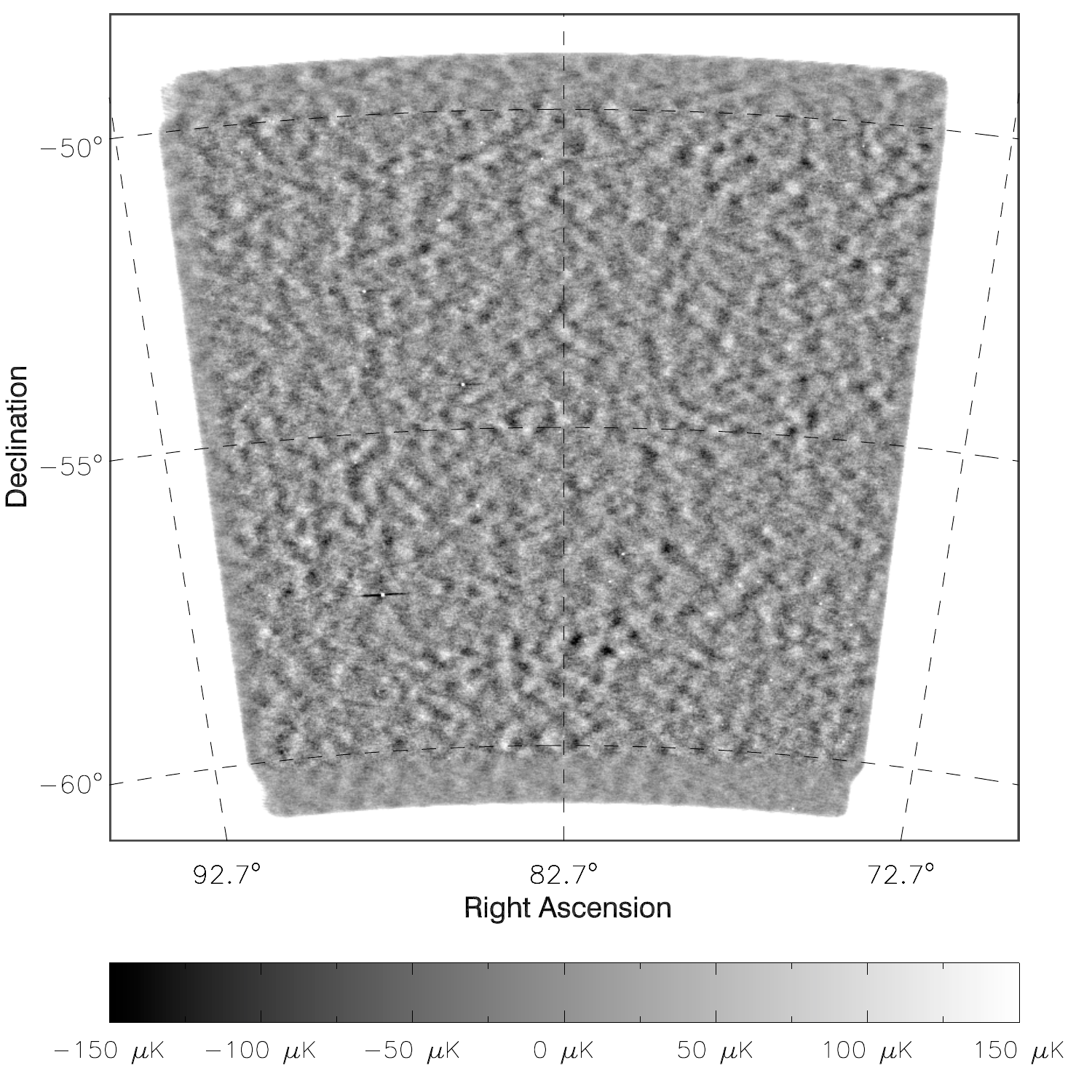}

\end{center}
\caption{Map of the \fivehs field at 220~GHz, with point sources
  unmasked during data filtering, using the oblique Lambert equal-area
  azimuthal projection.  This is the filter and projection scheme tailored for performing point-source analysis.  
  For display purposes, the map has been smoothed with a 1-arcmin FWHM Gaussian, 
  and high-noise regions near the boundary of the map have been masked.
{\label{fig:proj5_220}}}
\end{figure*}

\subsection{Weights}
\label{sec:weights}

The weights used to coadd data from many detectors into a
single-observation map are calculated by averaging the calibrated PSD
for each detector between 1 and 3~Hz.  
For the telescope scan speeds used in observations of the \fivehs
field, this frequency range corresponds to a multipole range of 
roughly $1500 < \ell < 4500$ in the scan direction, a reasonable overlap
with the scales of interest for source and cluster detection and for high-$\ell$
power spectrum analyses.  

When we average the data from
individual detectors to obtain a measurement of the map value in a
particular pixel, we calculate the total weight for that pixel
by combining the weights from each individual detector. These pixel
weights are used to combine the multiple single-observation maps into a final
coadded map (see Section \ref{sec:coadds}).  The
primary source of noise variation between single-observation maps is
the weather, which is accurately reflected in the
calibrated 1-3~Hz noise estimates for the individual detectors in all
but the poorest-weather days.

The weights for the final map are used to define the uniform-coverage
region.   We define the uniform-coverage
region as the area of the resulting map at 150~GHz for
which the weight in a map pixel 
exceeds 95\% of the median for 
all map pixels, resulting in an approximate area of 95~\degs.  

\subsection{Coadding Single-observation Maps}
\label{sec:coadds}

If all single-observation maps of the \fivehs field in a given observing 
band were identical up to noise variations captured accurately by 
the pixel weights described in the previous section, then the final
maps in each band would simply be weighted averages over all the
single-observation maps.  In reality, some fraction of single-observation
maps need to be excluded from the final coadd based on anomalous
behavior in the telescope, the weather, or the detectors.

Of the 421 total observations of the \fivehs field, we reject 55 from 
consideration for final maps because they were performed with atypical
gain settings for the detectors or had pointing control problems leading to
atypical coverage of the field.  We reject an additional 21
observations that were incomplete, due to the need to cycle the
cryogenic system.  

We examine the weights and the noise rms in the central 
five-by-five-degree~region of the individual observation maps and make 
further cuts based on the weight and noise rms statistics.  
We have found that two particular cuts are necessary to optimize the
final map noise and avoid biased signal values in the final map.
First, we cut any observations with anomalously high weights,
a condition which implies that the
recorded detector noise rms is lower than we can reasonably expect, 
possibly because changes in loading have significantly altered the 
detectors' operating point.  
We also cut individual maps in which the product of the median 
weight and the noise rms squared is anomalously high.
This can occur if the total map rms on all scales
does not track the 1-3~Hz noise on which the weights are based,
possibly due to anomalously poor weather.
The weights- and noise-rms-based cuts eliminate an additional  
20 individual maps at 150~GHz and 29 individual maps at 220~GHz.

All observations of the \fivehs field that survive the data selection cuts
were taken with the sun below the horizon.  
The moon was above the horizon for 52\% of these observations
but was never closer than 80 degrees from the field center.
Cross-checks performed for power spectrum
analyses using these data \citep{lueker10, shirokoff11, keisler11} have
shown no evidence of contamination from moon pick-up.
The observations of the \fivehs field 
sample the full range of 
azimuth directions.  Cross-checks performed for power spectrum
analyses using these data \citep{lueker10, shirokoff11, keisler11} have
shown no evidence of contamination from ground-based signals on
spatial scales that are measured in these maps.

After all of these checks and cuts, the final 150~GHz maps presented here are 
built from 321 individual observations totaling 621 hours, with a
median number of 304 individual detectors contributing data during
each observation.  For the 220~GHz maps, 313 observations totaling 605
hours are included, with a median number of 166 detectors contributing
data.

\section{Maps}
\label{sec:maps}
\subsection{Overview}
SPT maps
are constructed by a simple inverse-noise-weighted averaging of
observations of a given map pixel.  Noisy modes are filtered out of
the timestream rather than de-weighted in the mapmaking, and the
resulting map is not an unbiased estimate of the sky signal. An
understanding of both the noise and the effect of 
filtering on signal is essential for interpreting the maps.

  Ideally, the data processing
and map projection would be optimized separately for different science 
goals.  In order to present a relatively simple set of data products, we have
selected two variations on filtering and projection choices.  The
first set of maps, created with cluster-finding in mind, is produced
using point-source masking in the TOD filtering, using the
Sanson-Flamsteed projection.  The second set of maps, created with
point-source characterization in mind, is produced without masking
bright sources in the filtering, using the oblique Lambert
azimuthal equal-area projection.  

\begin{figure*}[ht]
\begin{center}
  \includegraphics[width=3in]{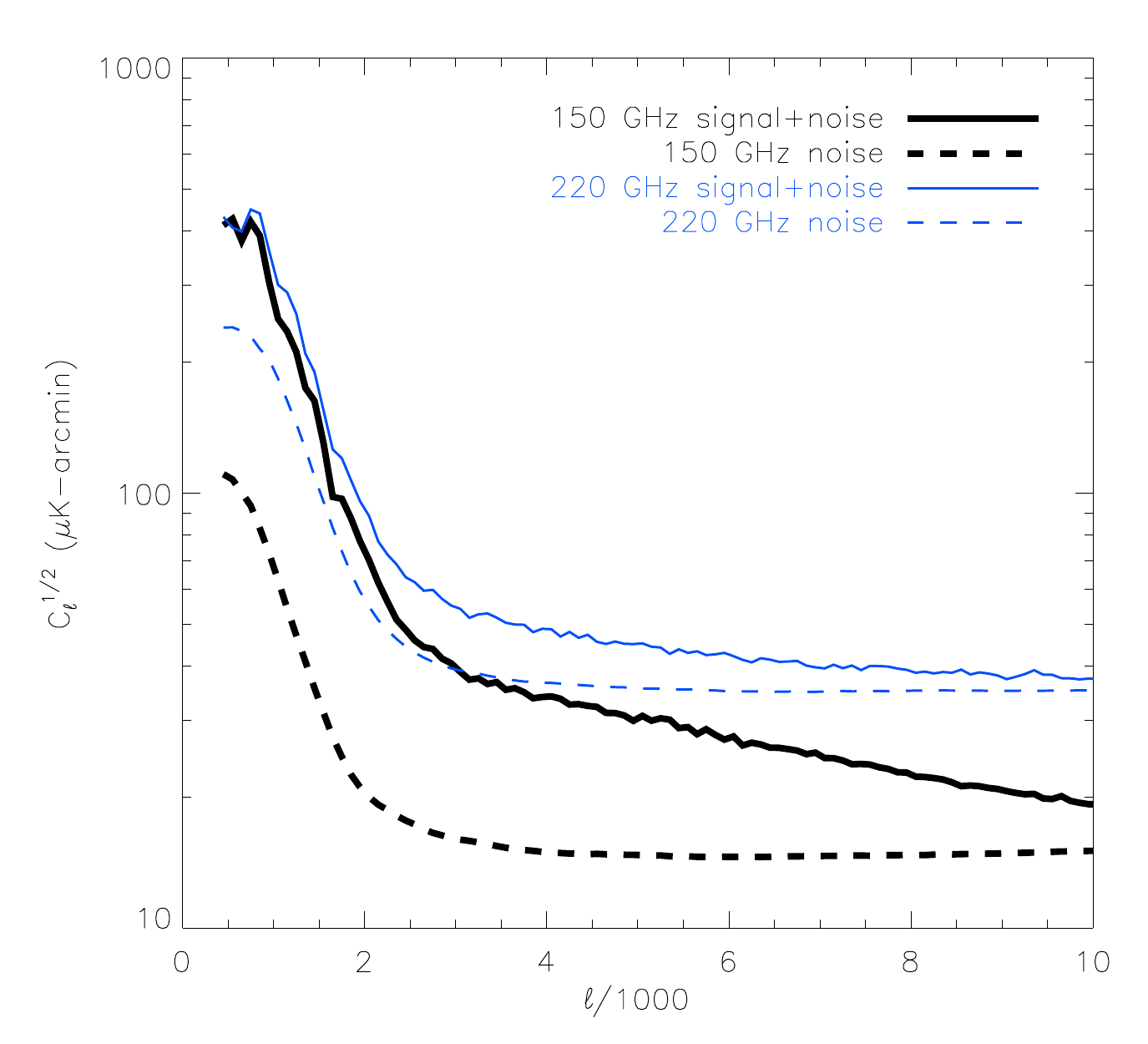}
  \includegraphics[width=3in]{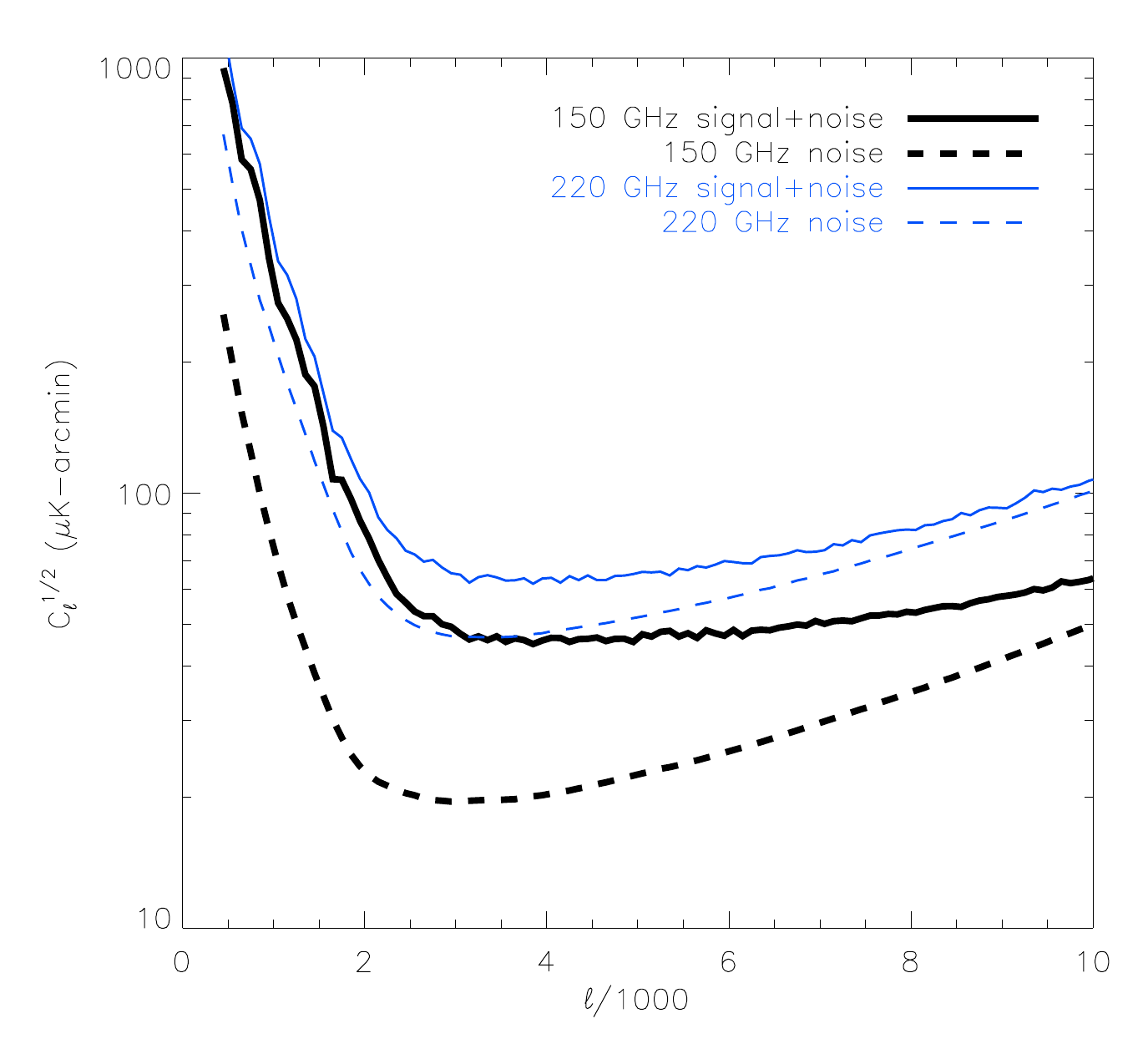}
\end{center}
\caption{\label{fig:signoise_1d} One-dimensional 
(azimuthally averaged in $\ell$ space) signal+noise and noise 
PSDs for each observing frequency.  
\textbf{Left Panel}: signal+noise and noise-only PSDs from
the raw map, uncorrected for beam and filtering effects.  
\textbf{Right Panel}: as in left panel, but with the beam and 
filter transfer function divided out.
Azimuthal averages are calculated
using rough noise weighting, with noisier modes at low $k_x$ (or $\ell_x$)
receiving less weight and modes below $k_x=\ell_x=400$ ignored entirely.
These demonstrate that the 150 GHz 
map is signal-dominated at nearly all spatial frequencies out
to $\ell=10000$, but that the 220~GHz map has significant noise 
contributions, particularly at $\ell \geq 5000$.  The essentially white character of 
the map noise is evident above $\ell=2000$ in the raw map PSDs.
The signal+noise PSDs with the beam and transfer function divided 
out show that the damping tail of the primary CMB (steeply falling with 
$\ell$) dominates the signal below $\ell \simeq 2500$, while the contribution
from point sources (flat with $\ell$) dominates above $\ell \simeq 2500$.}
\end{figure*}

The 150~GHz and 220~GHz maps with these processing and projection
choices are presented in Figures~\ref{fig:proj0_150} through
\ref{fig:proj5_220}.  For display, each map has been smoothed by convolving it with a 
1-arcmin FWHM Gaussian to suppress noise on scales smaller 
than the approximate size of the beam, and high-noise regions near 
the boundary of the map have been masked.  Figure \ref{fig:signoise_1d} shows
one-dimensional (azimuthally averaged in $\ell$ space) 
signal+noise and noise PSDs for each map (computed 
from the point-source-masked maps in the Sanson-Flamsteed projection).  
These plots are included to illustrate: 
1) where in spatial frequency the maps are signal- or noise-dominated; 
2) where the noise is white and where it has a ``red" spectrum due to atmosphere; 
and 3) what signals are contributing at different spatial frequencies.
In the following sections we discuss estimates
of the filtering and noise properties of the maps.

\subsection{Filter Transfer Functions}
\label{sec:tfs}

The beam and the timestream filtering are the two response functions
that have the most impact on the properties of sky signals recorded in
SPT maps.  While these
response functions are often combined, we separate them here because
they vary in different ways with the choice of map
projection.  The beam 
functions were described in Section~\ref{sec:beams}.  In this section,
we discuss how the time-domain filtering of the data from each scan
combines to affect the properties of signals in the
two-dimensional maps.  

The data filtering, as discussed in Section~\ref{sec:filtering},
involves low-pass and high-pass filters with a
wedge-average subtraction at each time sample.  These TOD
filtering operations combine with the coverage and projection
to result in filtering of signal on certain spatial scales
in the maps.  We characterize the effect of filtering on sky signals by
estimating the two dimensional filter transfer functions.  These are
the Fourier-domain functions representing the relative suppression of
signal power as a function of angular scale in the $x$ and $y$ dimensions
of the map. 

The transfer functions are typically estimated by
simulating observations of a known signal using the reconstructed
pointing for the real data, and passing
the simulated observations through the full data analysis pipeline.
Dividing the two-dimensional discrete FT of the
simulated map by the FT of the input signal yields an estimate of the filter
transfer function.  Assuming that the effect of filtering is linear,
the estimated transfer functions should be independent of the input
signal used in the simulations.  
In practice, the estimated transfer function depends very slightly on the 
input signal used, but the difference averaged over all spatial modes 
of interest is less than $1 \%$, even for input signals as drastically different 
as a point source and a simulated CMB sky.
For the maps presented here, we have
estimated the transfer functions using simulations of a Gaussian input
signal with an FWHM of 0.75 arcmin, chosen to probe the spatial
scales of interest in these maps.

\begin{figure*}[ht]
\begin{center}
\subfigure[Filter transfer function for 150~GHz]{\label{fig:proj0_tf_150} 
  \includegraphics[width=3in]{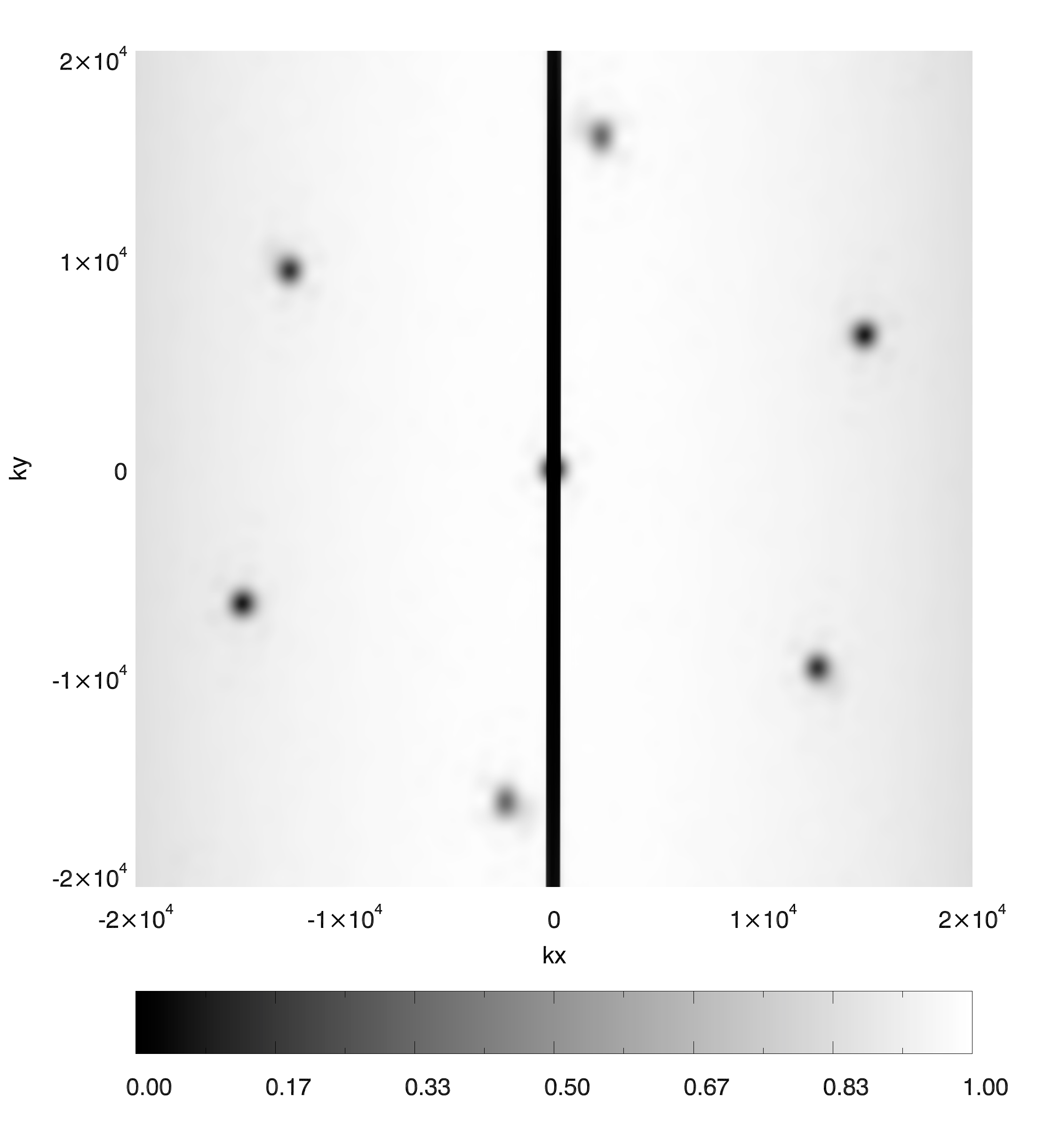}}
\subfigure[Filter transfer function for 220~GHz]{\label{fig:proj0_tf_220}
  \includegraphics[width=3in]{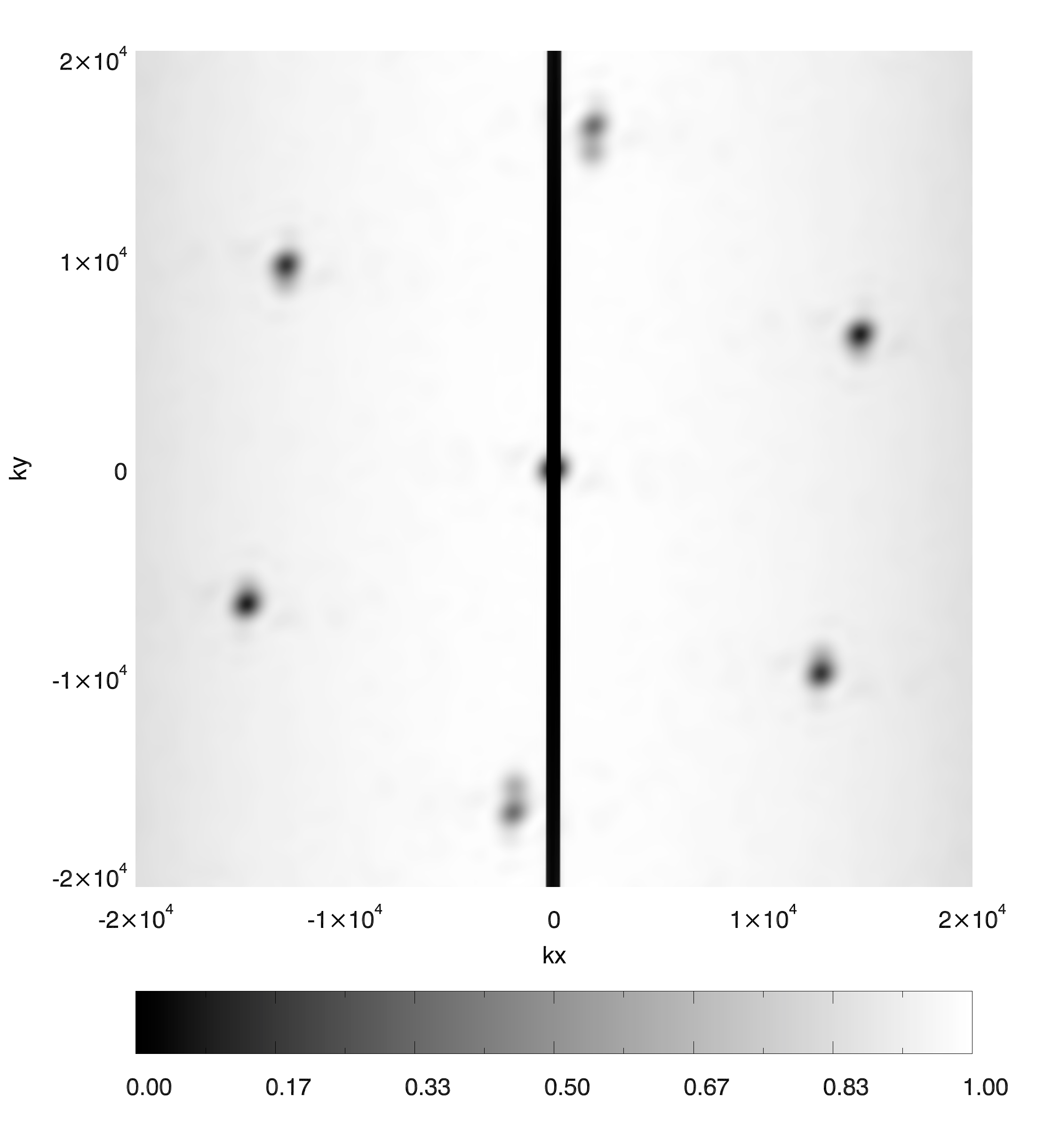}}
\end{center}
\caption{\label{fig:proj0_tf} Estimated two-dimensional transfer
  functions for the 150~GHz and 220~GHz maps employing point-source
  masking and the Sanson-Flamsteed projection.  These are averages
  over the entire map.  The dark stripe along $k_x=0$ is due to the 
  effective scan-direction high-pass filter from the polynomial and 
  Fourier-mode subtraction, the dark spot centered at $k_y=k_x=0$
  comes from the isotropic high-pass filter of the wedge mean subtraction,
  and the hexagonally spaced dark spots are due to the sensitivity of the
  wedge mean subtraction to modes with wavelengths and angles 
  corresponding to the detector array configuration.
  (See Sections \ref{sec:filtering} and \ref{sec:tfs} for details on
  filtering and associated transfer function features.)}

\end{figure*}

Figure
\ref{fig:proj0_tf} shows the estimated two-dimensional filter
transfer function for the maps tailored to cluster-finding (filtered with point sources masked and
presented with the Sanson-Flamsteed projection)
In the Sanson-Flamsteed projection, the telescope scan direction is
nearly equivalent to the $x$ direction in the map.  This means
that the low-pass and high-pass time-domain filters translate effectively into
low-pass and high-pass spatial filters along the $x$ direction.  
The wedge-average filtering acts as a roughly isotropic high-pass
filter.  In addition, it produces localized decrements in the transfer
functions at the 
spatial scales corresponding to the separation distances between
individual detectors.  Sky signal modes with wavelengths at
the detection separation scale and aligned in angle with the 
pixel configuration
will be almost
perfectly removed by the wedge-average subtraction.  The arrangement
of these features reflects the hexagonal symmetry of the detector
array.  The angle between the symmetry axis of the features and 
$k_x=0$ reflects the fact that the focal plane is slightly rotated 
with respect to the vertical axis of the telescope (to increase uniformity
of map coverage along the elevation direction).

 The
width of the reconstructed map varies as a function of declination in
the projection, so there is a slight 
change in the effective spatial cut-off scales for the high-pass and
low-pass filters at the top and bottom of the map.  
 Neglecting these small effects, the filtering is essentially
uniform across the maps.

\begin{figure*}[ht]
\begin{center}
\subfigure[Center filter transfer function for 150~GHz map]{\label{fig:tf_center_150_proj5} 
  \includegraphics[width=3in]{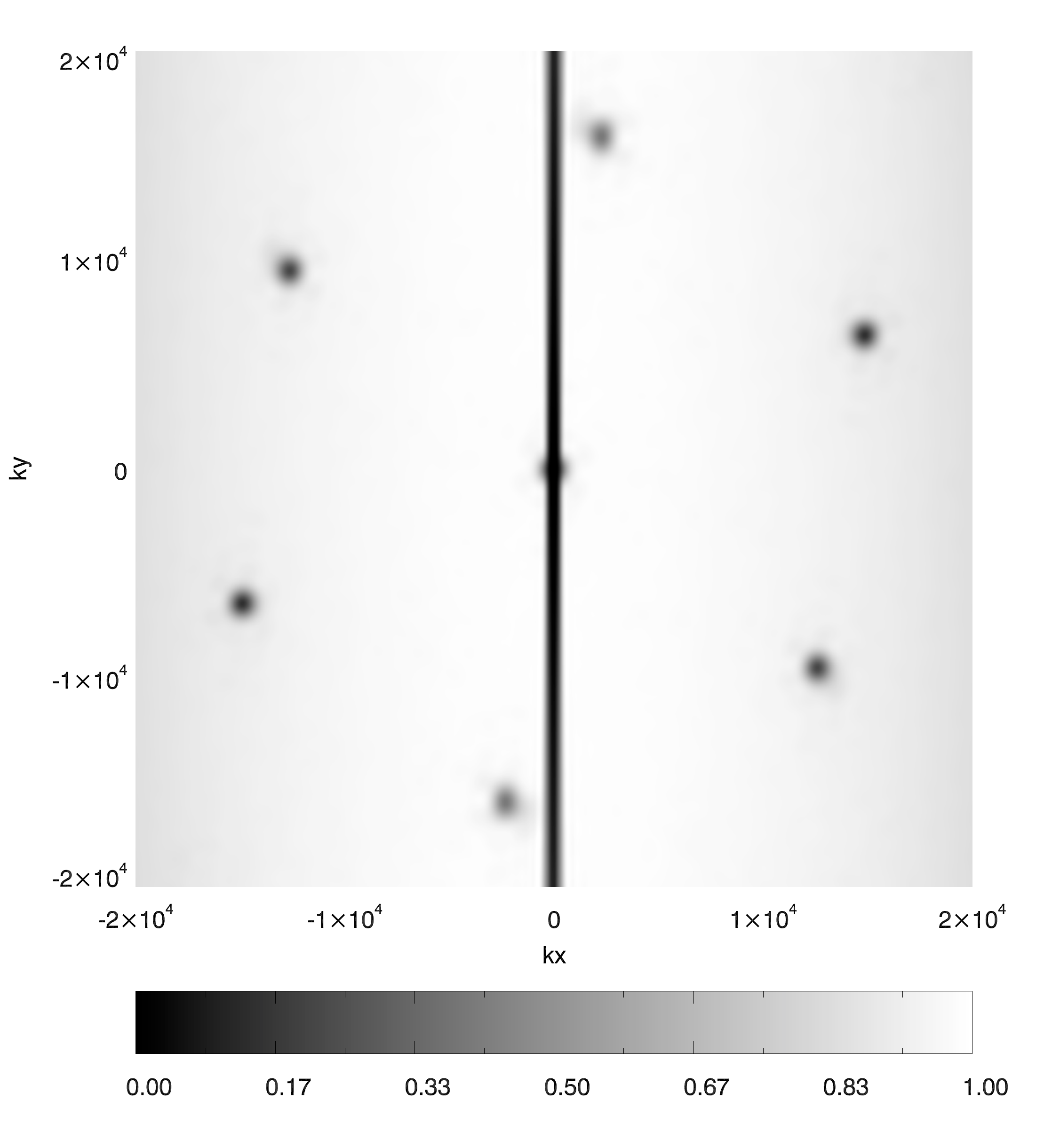}}
\subfigure[Corner filter transfer function for 150~GHz map]{\label{fig:tf_corner_150_proj5}
  \includegraphics[width=3in]{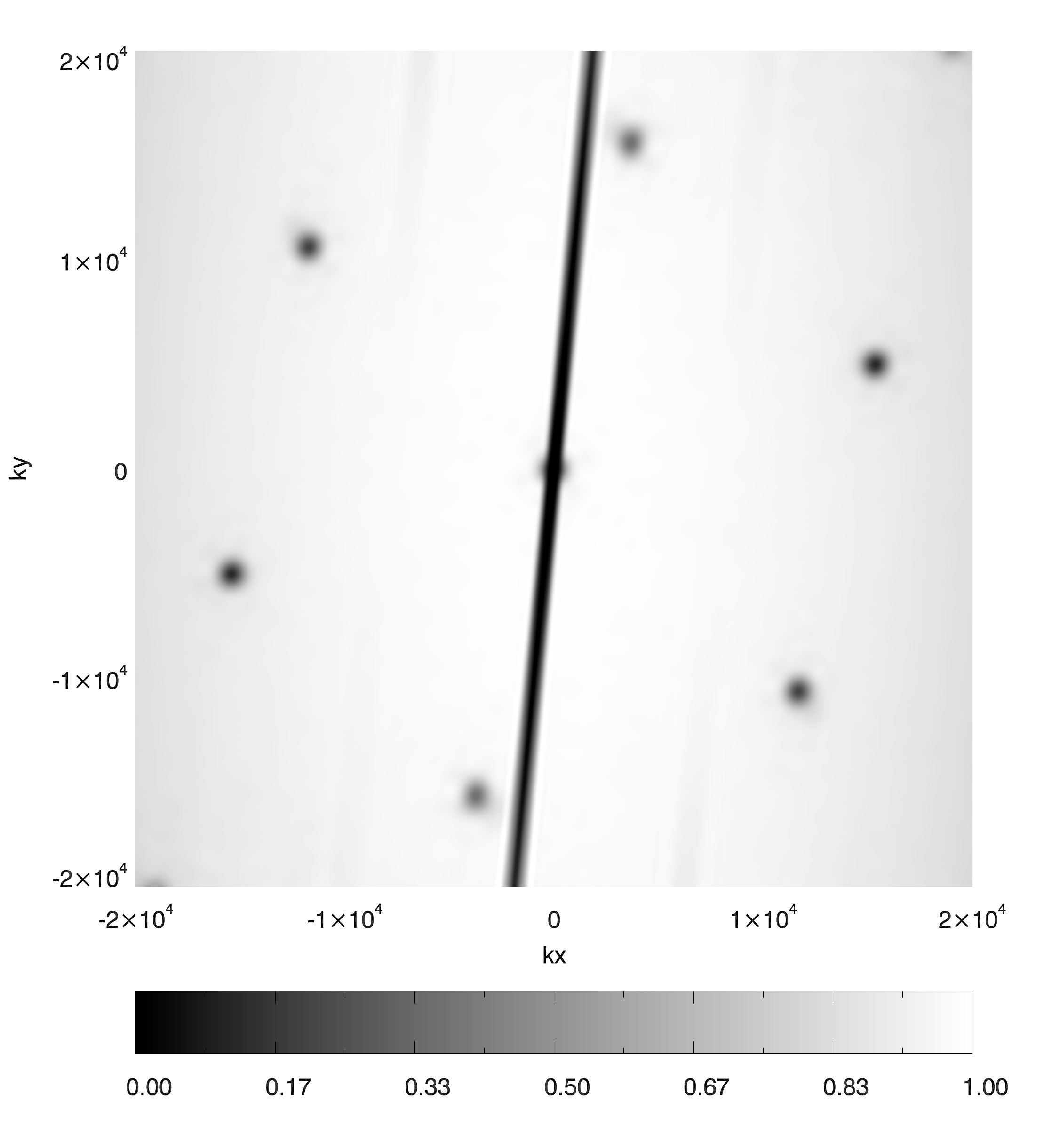}}
\subfigure[Center filter transfer function for 220~GHz map]{\label{fig:tf_center_220_proj5} 
  \includegraphics[width=3in]{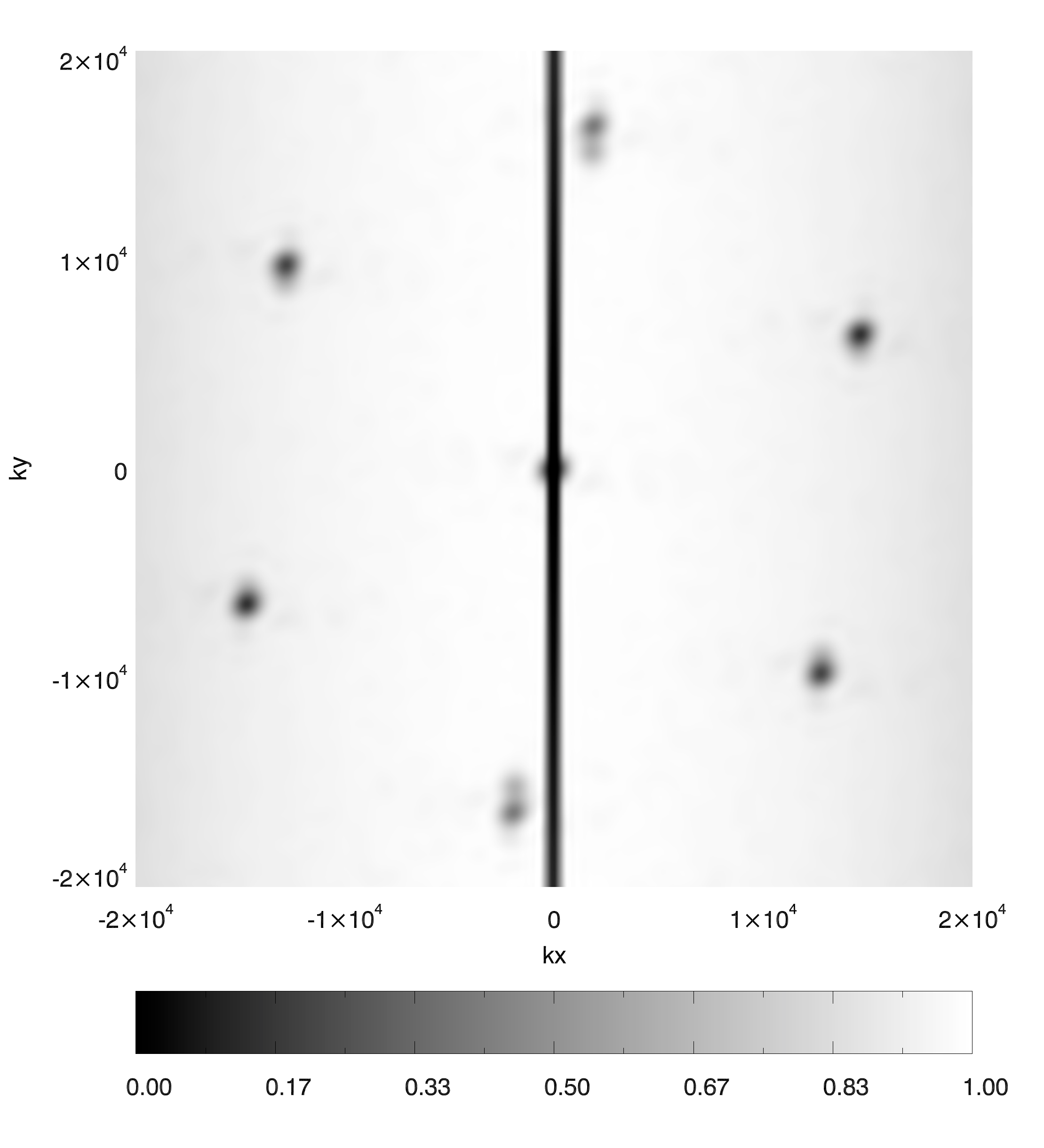}}
\subfigure[Corner filter transfer function for 220~GHz map]{\label{fig:tf_corner_220_proj5}
  \includegraphics[width=3in]{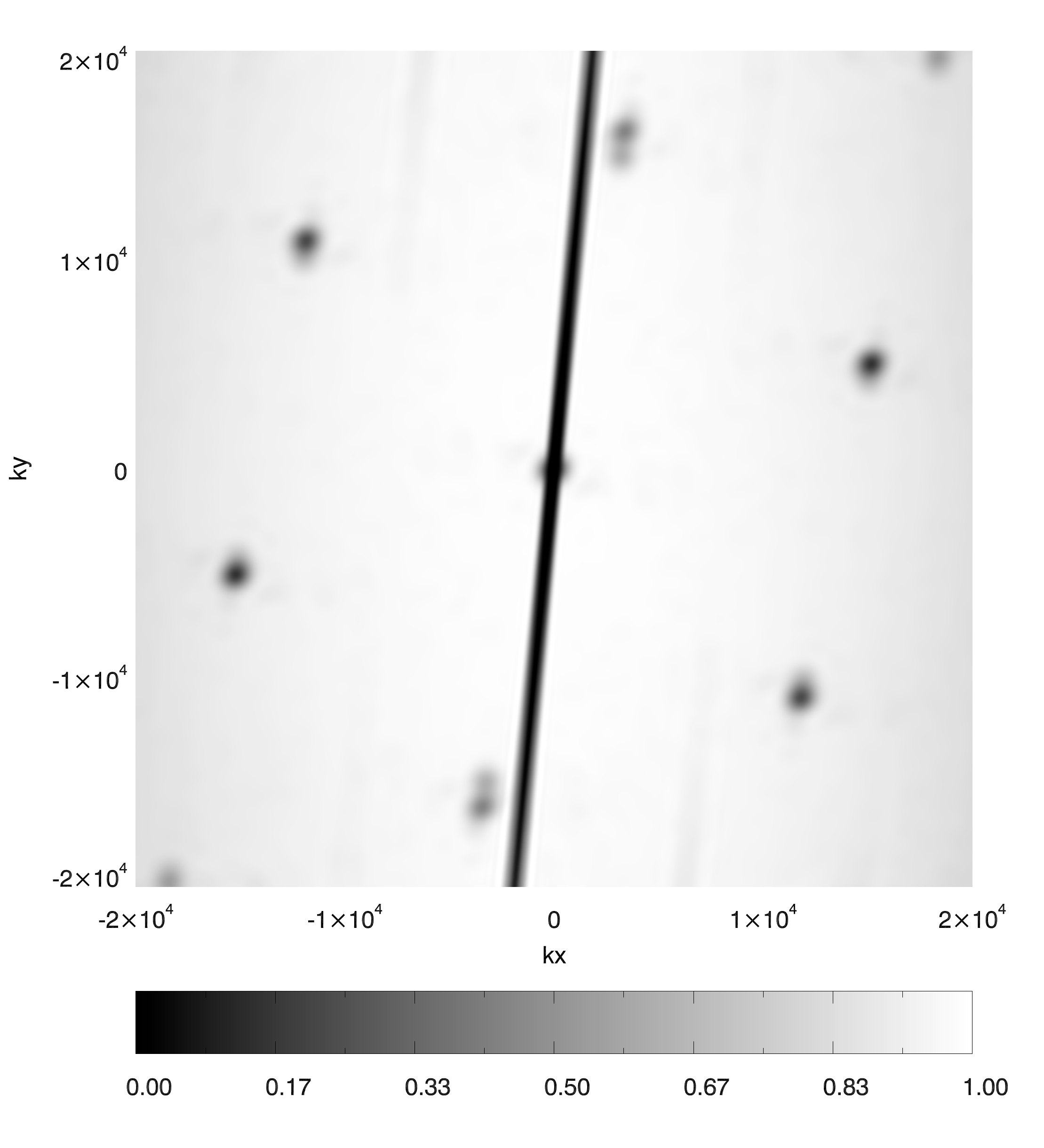}}

\end{center}
\caption{\label{fig:proj5_tf} Estimated two-dimensional filter transfer
  functions for the 150~GHz and 220~GHz maps employing no point-source
  masking and the oblique Lambert equal-area azimuthal projection.  The
  ``center'' filter transfer functions represent the innermost one-ninth
  of the map region, while the ``corner'' filter transfer functions represent
  one-ninth of 
  the map region located in the lower
  right corner. See the caption to Figure \ref{fig:proj0_tf} and Section \ref{sec:tfs} for a 
  discussion of the obvious features in these transfer functions. The rotation of the 
  features between center and corner transfer functions is discussed in Section 
  \ref{sec:tfs}.}

\end{figure*}

Filter transfer functions for the maps produced with the oblique
Lambert equal-area azimuthal projection are more complicated.  In this
projection, the telescope scans are at an angle to the $x$-direction in
the map, and this angle varies with position.  
This means that the two-dimensional transfer function rotates as a
function of map position (the angle of rotation is equal to the angle
between map rows and R.A., which is computed as a function of map
position in Equation \ref{eqn:rot}).
To address this rotation as well as more subtle
changes with position, we
have separately estimated the transfer functions with the input
Gaussian signal placed in nine different
locations in the map (the changes in transfer function across
the map are sufficiently slow and regular that nine locations
easily sample the full behavior).  Figure~\ref{fig:proj5_tf} compares the
estimated transfer function for the central portion of these maps to
the estimated transfer function in the lower right corner of the
maps.  The corner transfer function shows slight changes in the
effective spatial scale of the low-pass filtering, and
it exhibits additional diagonal features with amplitudes of 5-10\%.
These features arise because the filtered power associated with a
given bright source is spread diagonally across multiple rows of the
map.  In other words, they are a result of the interaction between the
filtering and the sampling function associated with the map
pixelization.  For
most applications, these features
can be neglected, as long as the rotation is taken into
account.  The transfer functions for all nine sub-regions of the map
are available online. 

\subsection{Map Noise}
\label{sec:psds}

\subsubsection{Map Noise Estimation and Description of Noise PSD Features} 
\label{sec:noisedesc}

Each individual two-hour observation of the field yields a map with
nearly identical sky signal contributions but independent atmospheric
and instrumental noise.  
We estimate 
the two-dimensional noise PSDs for the final coadded maps
using the jackknife noise estimation technique \citep{sayers09,halverson09}, 
which is used and described in many previous
SPT publications
\citep[e.g.,][]{staniszewski09,vanderlinde10,vieira10}.  Here, we
describe those functions, noting that we refer to the square root of
the noise power spectral density function as the PSD.

\begin{figure*}[ht]
\begin{center}
\subfigure[Noise PSD for the 150~GHz map]{\label{fig:psd_proj0_150} 
  \includegraphics[width=3in]{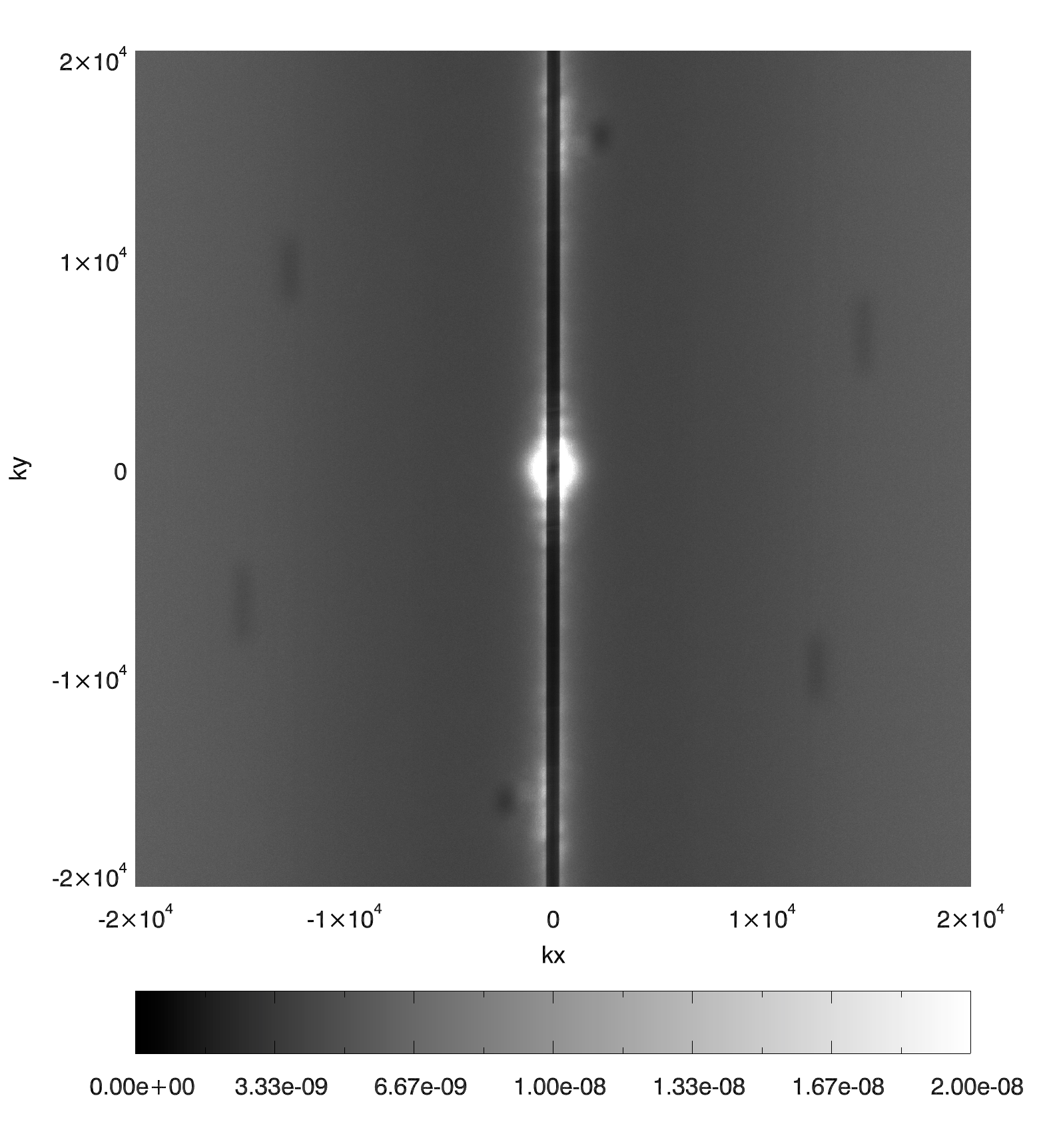}}
\subfigure[Noise PSD for the 220~GHz map]{\label{fig:psd_proj0_220}
  \includegraphics[width=3in]{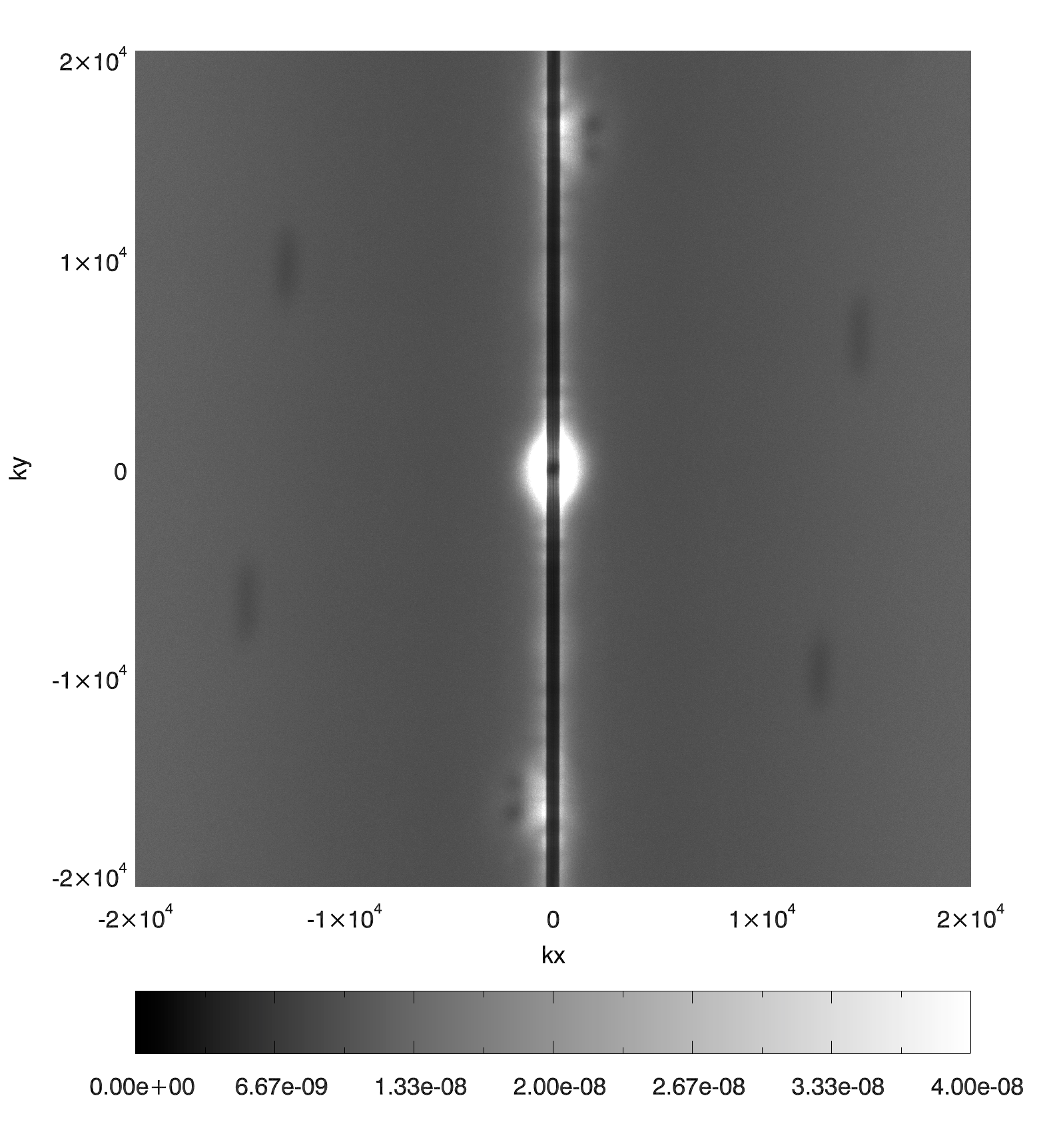}}
\end{center}
\caption{\label{fig:proj0_psds} Estimated two-dimensional noise PSDs
  for the 150~GHz and 220~GHz maps processed with point-source masking
  and the Sanson-Flamsteed projection.  Note that these are averaged
  over the whole map area. The high-noise area near $k_y=k_x=0$ is
  due to atmospheric noise, the uniformly increased noise at low $k_x$
  is due to low-temporal-frequency noise uncorrelated between detectors, 
  and the areas of excess noise at $k_y \simeq 15000$, $k_x \simeq 0$ are
  due to upmixed atmospheric power resulting from different sampling and 
  filtering of atmospheric noise in different detectors' time-ordered data.  
  As in the filter transfer functions (see Figures
  \ref{fig:proj0_tf} and \ref{fig:proj5_tf}), the dark stripe along $k_x=0$ is due to the 
  effective scan-direction high-pass filter from the polynomial and 
  Fourier-mode subtraction, 
  and the hexagonally spaced dark spots are due to the sensitivity of the
  wedge mean subtraction to modes with wavelengths and angles 
  corresponding to the detector array configuration.}

\end{figure*}

Figure~\ref{fig:proj0_psds} shows the averaged two-dimensional noise PSDs
estimated for the maps constructed with the Sanson-Flamsteed
projection.  For most of the signal region of interest for
cluster-finding and point-source characterization, the noise is
essentially white.
Because noise
power in the TOD translates to 
slightly different spatial scales at the top and bottom of the map,
there is an effective noise gradient of about 10\%, scaling with
the square root of the cosine of the declination in the map.  The PSDs
shown in Figure~\ref{fig:proj0_psds} and the depth values quoted in
Section~\ref{sec:mapaccount} are averages over all declinations in the map.

The most obvious non-white features in the PSDs are the concentration
of high noise at low $k$ (near the origin), the increase in noise toward
low $k_x$, and the noise cutoff at $k_x$ values below $\sim 300$.  The 
increasing noise at low $k_x$ is due to $1/f$ noise that is
uncorrelated between detector channels.  The cutoff at low $k_x$ is due to the
high-pass filter, just as in the filter transfer functions in Section
\ref{sec:tfs}.  High noise near the origin is due to brightness fluctuations in the atmosphere.
These fluctuations 
are spatially isotropic
and have a very ``red'' spectrum, with most of the power on large spatial
scales.  Depending on the scan speed of the instrument, in any given 
observation the atmosphere will either be imaged as if it were a stationary
sky signal---in which case the atmosphere will appear as isotropic 
noise at low $k$, just as it does in the composite PSDs shown here---or 
it will blow past the array in the direction of the wind---in which case it 
will appear elongated in $k$-space along the wind direction in
a single-observation PSD.  Even in the latter case, changes in wind 
direction over the course of many observations will tend to make the
atmospheric contribution essentially isotropic.

The visible decrements in the PSDs in Figure~\ref{fig:proj0_psds}
correspond to the 
features previously described in the two-dimensional transfer
functions.
Residual atmospheric noise 
contributes excess noise at large spatial scales, but the majority of
the atmospheric and instrumental noise is removed by the high-pass
filter.  A small amount of noise power
``leaks'' into the central high-pass-filtered region of the Fourier
plane because of the subsequent application of the wedge-average
filtering.  The areas of excess noise at high $k_y$/low $k_x$ are
explained as ``upmixed'' large-scale noise power.  This upmixing
occurs because the sampling and filtering of large-scale atmospheric 
features is slightly different in the TOD of each individual detector.

\begin{figure*}[ht]`
\begin{center}
\subfigure[Center noise PSD for the 150~GHz map]{\label{fig:psd_center_proj5_150} 
  \includegraphics[width=3in]{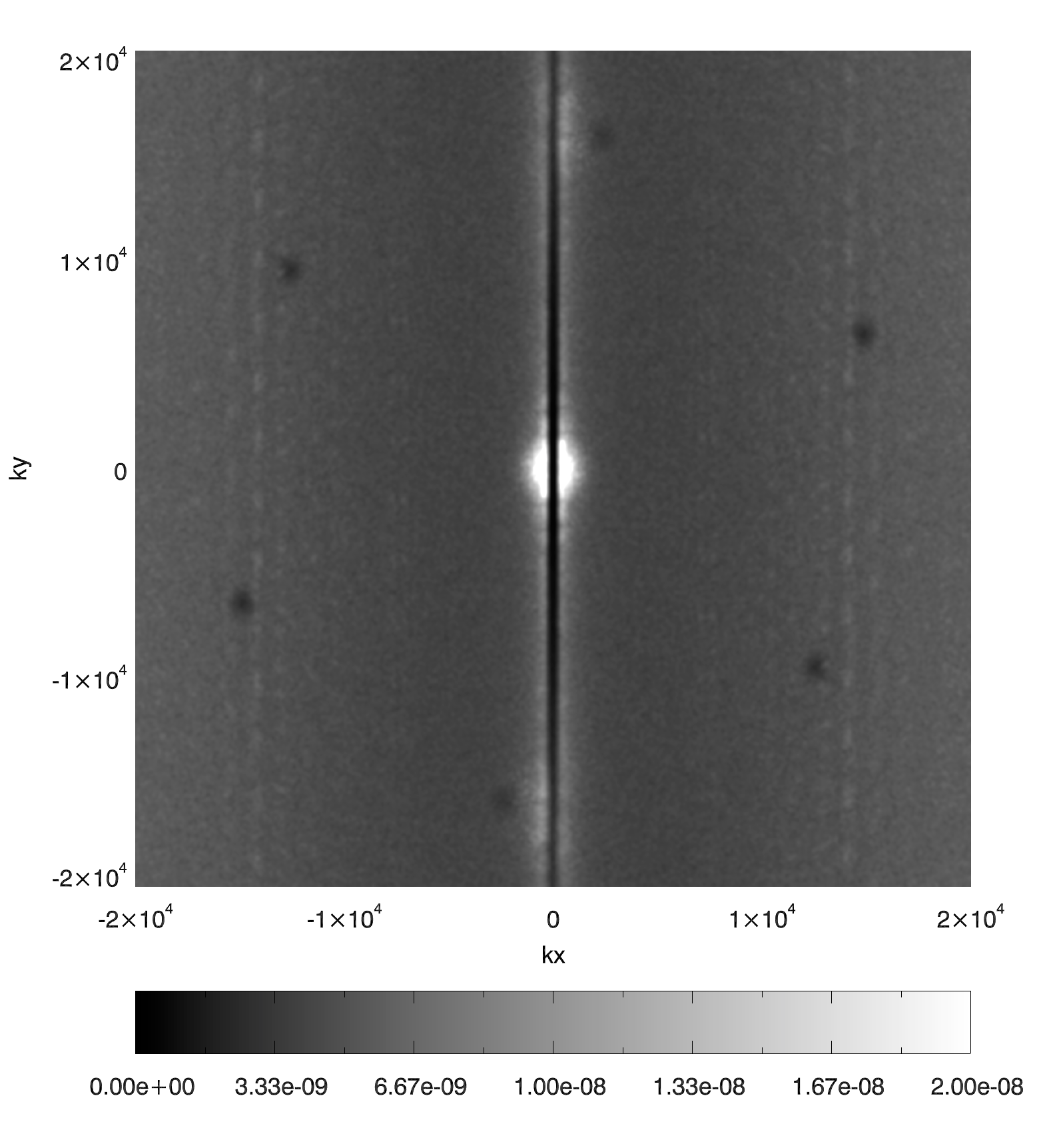}}
\subfigure[Corner noise PSD for the 150~GHz map]{\label{fig:psd_corner_proj5_150}
  \includegraphics[width=3in]{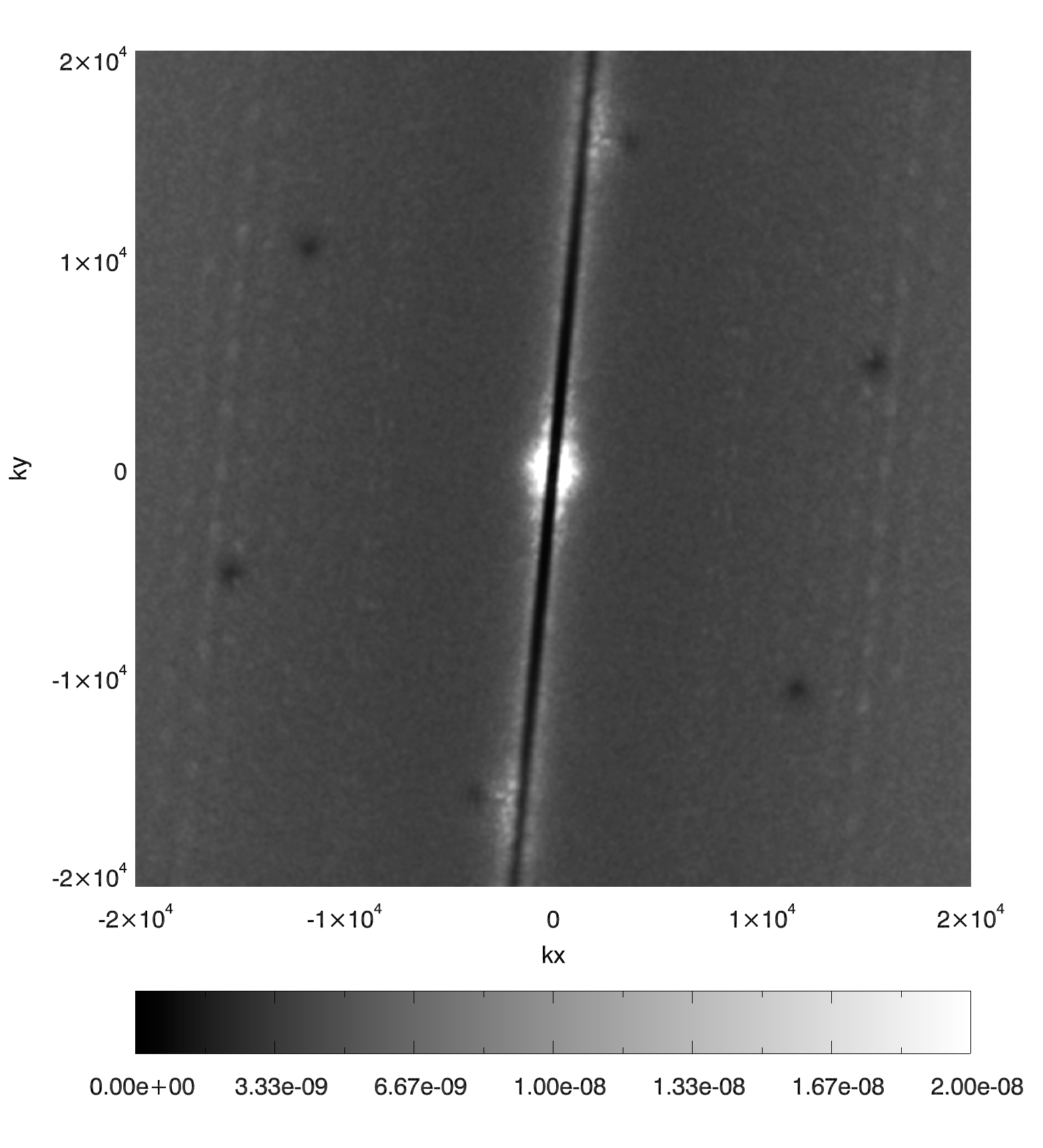}}
\subfigure[Center noise PSD for the 220~GHz map]{\label{fig:psd_center_proj5_220} 
  \includegraphics[width=3in]{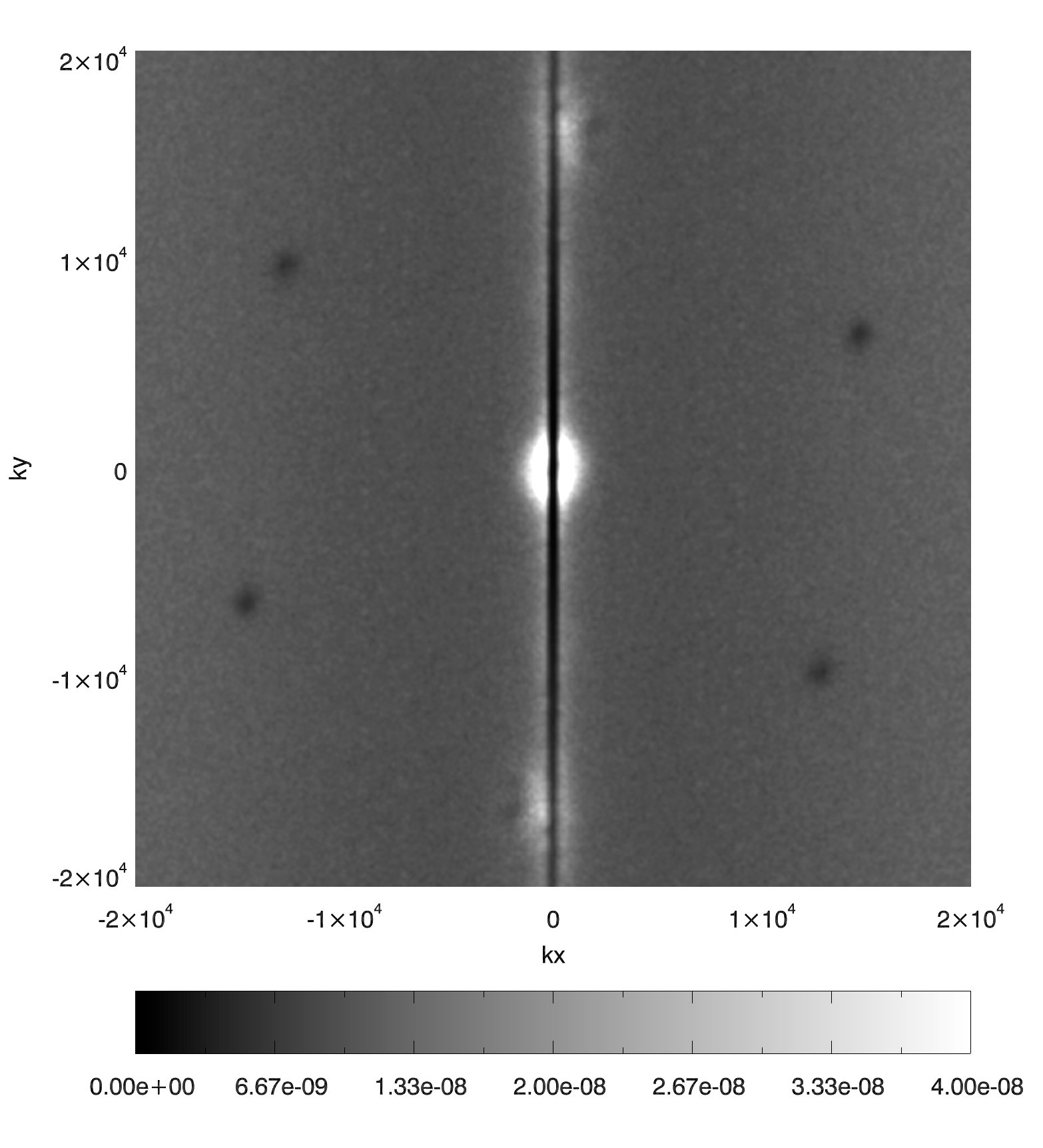}}
\subfigure[Corner noise PSD for the 220~GHz map]{\label{fig:psd_corner_proj5_220}
  \includegraphics[width=3in]{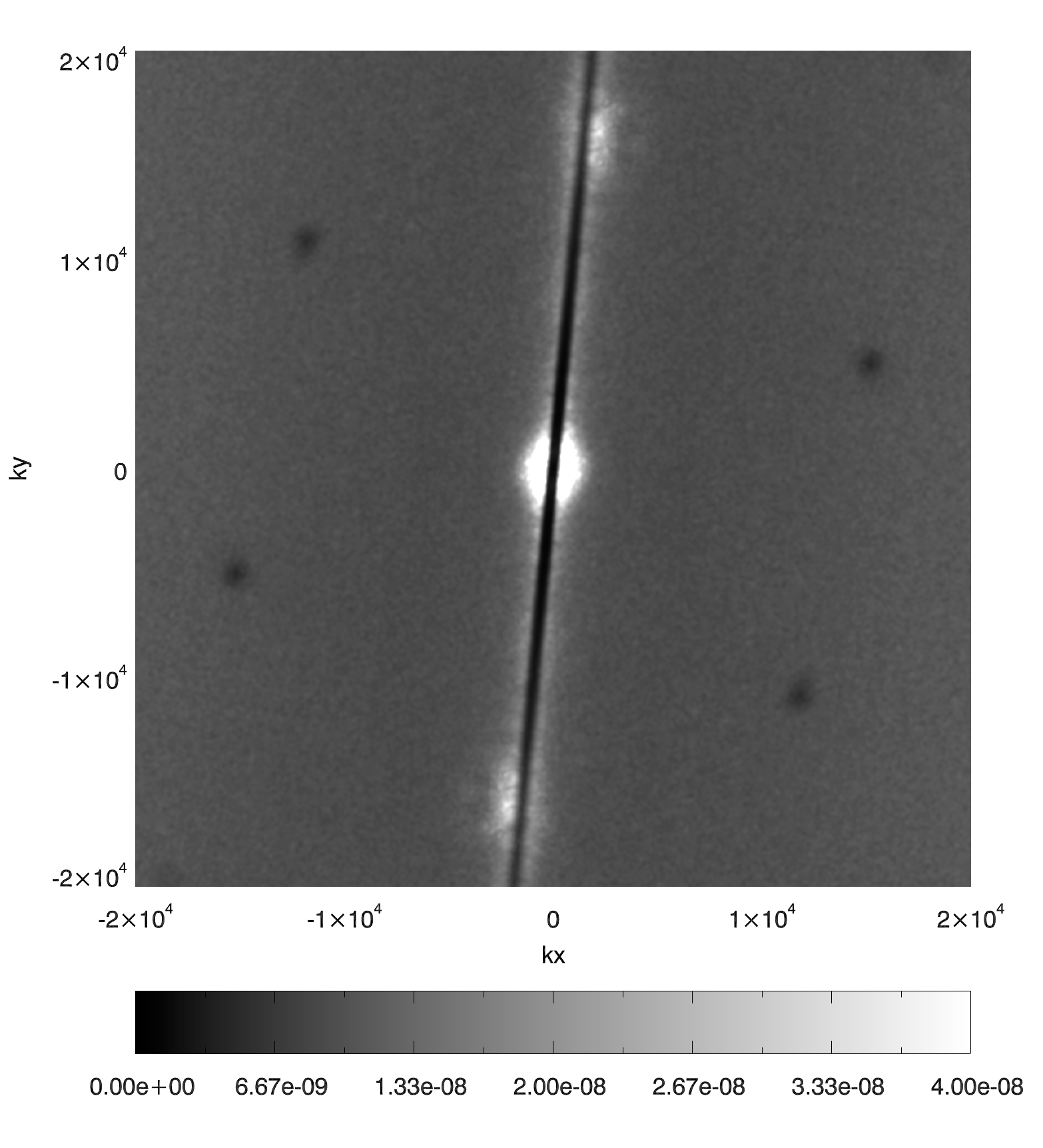}}
\end{center}
\caption{\label{fig:proj5_psds} Estimated two-dimensional noise PSDs
  for selected sub-regions of the 150~GHz and 220~GHz maps processed
  with no point-source masking and the oblique Lambert equal-area
  azimuthal projection.  The ``center'' PSDs represent the innermost one-ninth
  of the map region, while the ``corner'' PSDs represent one-ninth of
  the map region located in the lower
  right corner. See the caption to Figure \ref{fig:proj0_psds} and Section \ref{sec:noisedesc} for a 
  discussion of the obvious features in these PSDs. The faint lines at constant $k_x$
  (not apparent in the Figure \ref{fig:proj0_psds} PSDs) are due to readout lines (see
  Section \ref{sec:noisedesc} for details). The rotation of the 
  features between center and corner PSDs is discussed in Section 
  \ref{sec:noisedesc}.
}

\end{figure*}

Figure~\ref{fig:proj5_psds} shows the averaged two-dimensional noise
PSDs estimated for the center and lower right corner of the maps with
the oblique Lambert equal-area azimuthal projection.  The properties
are similar to those for the Sanson-Flamsteed projection, with the
added rotation for positions away from the center of the map (the reason
for this rotation and the value of the rotation angle are identical to those 
discussed for the filter transfer functions in Section~\ref{sec:tfs}).  In
these PSDs, the effect of residual high-frequency readout-related line
features is visible as vertical strips at high $k_x$.  These are less
visible in the full-map average PSDs presented in
Figure~\ref{fig:proj0_psds}, but the features are present.   These features
have a negligible effect on analyses performed with the maps.

As with
the transfer functions, PSDs have been estimated for nine sub-regions
of the oblique Lambert equal-area azimuthal projection maps, and these data
products are available to download. 

\subsubsection{Map Noise Accounting}
\label{sec:mapaccount}

The noise PSDs discussed in Section~\ref{sec:noisedesc} can be used to estimate
the effective map depths, but it is important to highlight that these
numbers depend on the absolute calibration and on the normalization of the beam
functions.  In previous SPT publications, map depth has been quoted 
under the approximation that the SPT beams are Gaussian, with 
$B(\ell)=1$ at $\ell=0$.  Using the measured SPT beam presented here, 
with $B(\ell)$ normalized to unity at $\ell=800$, the quoted map depth numbers 
are slightly different, but, as we will show, the two sets of beam/calibration/depth
numbers are consistent with each other.  Put another way, the 
fundamental quantity of interest for science is the 
noise after the beam and transfer function have been deconvolved, and the 
deconvolved noise we measure here is 
consistent with the Gaussian-beam estimates in previous papers, as shown below.

Quadrature-averaging an annulus of the PSDs between $\ell = 4000$ and
$\ell=5000$, we estimate a depth of 
\noiseonftybeamcal~$\mu$K-arcmin at 150~GHz and 
\noisettwntybeamcal~$\mu$K-arcmin at 220~GHz.  For this
estimate, 
as in Figure \ref{fig:signoise_1d}, azimuthal averages are calculated
using rough noise weighting, with noisier modes at low $k_x$ (or $\ell_x$)
receiving less weight and modes below $k_x=\ell_x=400$ ignored entirely.
This estimate
of map depth must be interpreted using the beams reported in this
release, because the calibration applied to the maps depends on the
beam function shape and normalization.  

For simple calculations of signal sensitivity, it is common to
use a Gaussian approximation to the beam.  If we approximate the beams
as Gaussians with FWHM 1.15 arcmin and 1.05 arcmin for the
150~GHz and 220~GHz maps, respectively, we also need to modify the
effective calibration of the maps.  Within the $\ell$ range used for
estimating the map depths, the average ratio of the true beam
function to the Gaussian beam function can be used to adjust the depth
figures.  Over the $\ell$ range used in this noise estimate, that 
ratio is \gaussratonfty \ at 150~GHz and \gaussratttwnty \ at 220~GHz.
The depths appropriate for a Gaussian approximation to the
beam are thus approximately 
\noiseonftygauss~$\mu$K-arcmin at 150~GHz and
\noisettwntygauss~$\mu$K-arcmin at 220~GHz, as quoted in the abstract.
These numbers are consistent with the quoted values of 
18~$\mu$K-arcmin and 40~$\mu$K-arcmin from previous SPT publications
\citep[e.g.,][]{vanderlinde10,shirokoff11}.

The 150~GHz beam-deconvolved noise at $\ell=3000$ 
(see Figure \ref{fig:signoise_1d}, right panel) is 
\noiseonftydecon~$\mu$K-arcmin.
This is equal to the beam-deconvolved noise amplitude
at this multipole for a 1.15-arcmin FWHM Gaussian beam and 
\noiseonftydecongauss~$\mu$K-arcmin white noise.
The 220~GHz beam-deconvolved noise at $\ell=3000$ 
(see Figure \ref{fig:signoise_1d}, right panel) is 
\noisettwntydecon~$\mu$K-arcmin.
This is equal to the beam-deconvolved noise 
at this multipole for a 1.05-arcmin FWHM Gaussian beam and 
\noisettwntydecongauss~$\mu$K-arcmin white noise.

In Section~\ref{sec:instrument}, we reported mapping speeds for the 
full detector array at 150 and 220~GHz, using the absolute
temperature calibration described in this release.  These numbers can
be combined with the total integration time on the \fivehs field reported in
Section~\ref{sec:observations} to obtain an estimate of the
expected depth of the maps.  A simple calculation using the numbers
for mapping speed, field size, and integration time
to determine depth would yield expected depths of roughly 
\noiseonftypredraw \ and
\noisettwntypredraw~$\mu$K-arcmin at 150 and 220~GHz, respectively.  These
numbers are significantly lower
than the 
\noiseonftybeamcal \ and
\noisettwntybeamcal~$\mu$K-arcmin calculated directly from the noise
PSDs.  However, several factors reduce the time spent on the
uniform-coverage region for which the field size of $95$~\degss \  is
reported.  First, with a one-degree-wide detector array we must map 
an 11-by-11-degree region to end up with a 10-by-10-degree region
which is seen by every bolometer.  This results in an efficiency loss 
of roughly $20\%$.  Second, the 
constant-velocity, bolometer, and scan cuts described in Section \ref{sec:preprocessing} result 
in an approximately $15\%$ loss.  
When these cuts and inefficiencies
are taken into account, the expected depths are 
\noiseonftypredeff \ and \noisettwntypredeff~$\mu$K-arcmin 
at 150 and 220~GHz, within 10$\%$ of the observed map depth.  All map
depth estimates assume that the noise is white, which is known to be
only an approximation even for the $\ell$ range used for these estimates.

\subsection{Comparison to Past SPT Analyses}
\label{sec:catalogs}

The maps and data products presented in this release are very similar, 
but not identical,
to maps of the \fivehs field used by the SPT team in previously published
scientific results.  These published results include catalogs of 
SZ-selected galaxy clusters \citep{staniszewski09, vanderlinde10,williamson11} 
and emissive sources \citep{vieira10}, and measurements of the
CMB angular power spectrum \citep{lueker10,shirokoff11,keisler11}.  Minor variations in data
selection, filtering, map projection, and approaches to defining transfer functions and
noise PSDs are expected to cause small changes in catalogs of
objects extracted from these maps and to estimates of the CMB power spectrum.  
We have reproduced two past catalogs and two past power spectrum analyses
using this map release 
to verify that the differences are negligibly small.

Using the 150~GHz map from this release that was produced with
the Sanson-Flamsteed projection and point-source masking, we have reproduced
the analysis reported in \citet{vanderlinde10}.  All \fivehs clusters
presented in \citealt{vanderlinde10} (10 out of 21 total clusters in that work, 
which also included data from the \twentythreehs field) are clearly 
detected in our analysis with the data products from this release.
The signal-to-noise values for these 10 clusters in our analysis 
agree with the values quoted in \citet{vanderlinde10} to within 
$2 \%$ in the mean, with an rms scatter of  $4 \%$.  This
check was performed using the estimated PSD presented in this release, but with
the analytic transfer function estimate described in
\citet{vanderlinde10}.  A cross-check using the simulated transfer functions and
full two-dimensional beam functions presented in this release
produces identical results at the sub-percent level.

Using both the 150~GHz and 220~GHz maps from this release that were
produced with the oblique Lambert equal-area azimuthal projection and no
point-source masking, we have reproduced the analysis reported in
\citet{vieira10}.  This check was performed using the
noise PSDs, transfer functions, and beams presented in this release.  
The \citet{vieira10} catalog is faithfully reproduced using the
data products in this release, with every $5 \sigma$ source from \citet{vieira10}
detected at $\ge 4.4 \sigma$ in our analysis and mean signal-to-noise 
within $0.1 \%$ of the \citet{vieira10} values at 150~GHz and $1.0 \%$ at 220~GHz.  
The recovered fluxes of the sources in this analysis 
compared to the \citet{vieira10} fluxes are $3.1 \%$ lower at 
150~GHz and $7.5 \%$ lower at 220~GHz.
These differences are due to improved  
beam and calibration estimates.  If we repeat the analysis using the
maps, filter transfer functions, and noise PSDs presented here and
the beam and calibration used in the \citet{vieira10} analysis, 
we reproduce the 150 and 220~GHz flux numbers to better than $1 \%$.

Finally, we construct a CMB power spectrum estimate from the 150~GHz data
and compare it to the power
spectrum of this field as measured in  \citet{shirokoff11} and \citet{keisler11}.  
In each comparison, we use the calibration used in the 
\citet{shirokoff11} or \citet{keisler11} analyses.  
We construct the power spectrum by Fourier transforming the
point-source-masked 150~GHz map times an adapted version of the apodization
mask used in \citet{shirokoff11} or \citet{keisler11}, correcting for the beam and
transfer function, squaring, azimuthally averaging in $\ell$ space, and subtracting 
off a noise bias calculated by azimuthally averaging the squared noise PSD 
(also corrected for beam and transfer function).  Fourier modes with significantly 
elevated noise or where the transfer function is near zero were excluded from the
signal and noise averages.  

We find that this simply calculated power spectrum of the \fivehs map presented here,  
using the \citet{shirokoff11} calibration and masks, agrees with
the power spectrum of this field from the \citet{shirokoff11} analysis to within 
$3\%$ in power ($1.5\%$ in temperature).  This small difference is attributable
to mode-mixing
and window-function effects that are not taken into account in the simple analysis
performed here.  
We find that the power spectrum of the \fivehs map presented here, calculated 
using the \citet{keisler11} calibration and masks, agrees with
the power spectrum of this field from the \citet{keisler11} analysis to within 
$1\%$ in power ($0.5\%$ in temperature).

\section{Conclusions and Available Data Products}
\label{sec:conclusions}

We have presented the
first publicly released maps from the South Pole Telescope, of a 95~\degss
field observed during the 2008 season in two frequency bands (150 and 
220~GHz).  We have
described the observations, data selection, filtering, and mapmaking
approaches used to create these maps.  In addition, we have characterized
the instrument bands and beams, the filter transfer functions, and the noise
properties of the maps.  The maps and auxiliary data products documented in this paper
are available online for download and use by the broader community. The full set of data products, available at 
http://pole.uchicago.edu/public/data/maps/ra5h30dec-55
and from the NASA Legacy Archive for Microwave Background Data Analysis
server, includes:

\begin{itemize}

\item Maps:  Two versions of both the 150~GHz and 220~GHz maps of the
  \fivehs field are given.  One version uses the Sanson-Flamsteed
  projection and point-source masking in the filtering, and the other
  version uses the oblique Lambert equal-area azimuthal projection with no
  point-source masking.  

\item Beams:  Two-dimensional beam functions for 150~GHz and 220~GHz
  are provided, as well as one-dimensional beam averages in real and
  Fourier-space representations and an estimate of beam uncertainties.  

\item Bands:  Measured band-pass functions for 150~GHz and 220~GHz are
  provided.

\item Filter Transfer Functions:  Two sets of transfer functions are
  provided, corresponding to the two choices of projection and
  filtering.  For the oblique Lambert equal-area azimuthal projection,
  transfer functions estimated for 9 sub-regions of the map are
  given.

\item Noise PSDs:  Two sets of noise PSDs are provided, corresponding
  to the two choices of projection and filtering.  For the oblique Lambert
  equal-area azimuthal projection, PSDs estimated for 9 sub-regions of
  the map are given.

\end{itemize}

An example of a calculation using these data products is described in Appendix 
\ref{app:examplecalc}.

 We have checked
that these data products, when used in analyses similar to 
published analyses of this field's data, faithfully reproduce the published 
results.  The data release and this accompanying paper are the first step 
toward an eventual release of data from the full 2500~\degs, 
three-band SPT-SZ survey.

\begin{acknowledgments}
The South Pole Telescope is supported by the National Science
Foundation through grants ANT-0638937 and ANT-0130612.  Partial
support is also provided by the NSF Physics Frontier Center grant
PHY-0114422 to the Kavli Institute of Cosmological Physics at the
University of Chicago, the Kavli Foundation, and the Gordon and Betty
Moore Foundation.  
The McGill group acknowledges funding from the National   
Sciences and Engineering Research Council of Canada, 
Canada Research Chairs program, and 
the Canadian Institute for Advanced Research. 
Partial support at Harvard is provided by NSF grants AST-1009012,
AST-1009649, and MRI-0723073.
B.A. Benson is supported by a KICP Fellowship.
M. Dobbs and N. Halverson acknowledge support from Alfred P. Sloan Research Fellowships.
R. Keisler acknowledges support from NASA Hubble Fellowship grant HF-51275.01.
J. Mohr acknowledges support from the DFG supported Excellence Cluster
Universe and the transregio program TR33: Dark Universe.
B. Stalder acknowledges partial support from the Brinson Foundation.
We acknowledge the use of the Legacy Archive for Microwave Background Data Analysis (LAMBDA). 
Support for LAMBDA is provided by the NASA Office of Space Science.
\end{acknowledgments}

\appendix
\section{Example calculation}
\label{app:examplecalc}

As a simple introduction to the use of the data products presented here, and 
as a means for users to check that they are interpreting the data as intended, 
we present an example of a typical calculation, namely the estimation of the 
flux of a point source in the map.

The most straightforward way to estimate the flux of a point source 
with a known position is to read off the map value at the pixel that corresponds to 
that position, and to convert that value in K-CMB to a value in MJy/sr using the conversion 
factors in Table \ref{tab:bands}.  To convert that to a point source flux in Jy, one
needs an effective solid angle.  It is tempting to simply use the solid angle of the 
beam; however, that implicitly assumes
that signals in the map have undergone no filtering beyond 
beam smoothing, which is not necessarily the case.  In the case of a filtered map, the effective
solid angle to use in this calculation is \citep[e.g.,][]{vieira10}
\begin{equation}
\Delta \Omega = \left [ \frac{1}{4 \pi^2} \int d^2 k \
B(k_x,k_y) \ F(k_x,k_y) \right ]^{-1},
\end{equation}
where $B(k_x,k_y)$ is the two-dimensional Fourier-space beam function and 
$F(k_x,k_y)$ is the two-dimensional Fourier-space filter transfer function.

This calculation will give an unbiased estimate of the source flux, but the estimate
will be noisier than it has to be, because all spatial modes are treated equally, regardless of noise.
If the map is filtered further to downweight noisy modes, the signal-to-noise on 
objects with a known angular profile (such as point sources) can be improved.
The Fourier-space optimal filter for sources of a known shape is \citep[e.g.,][]{haehnelt96}
\begin{equation}
\psi(k_x,k_y) \equiv \frac{\tau(k_x,k_y) N^{-1}(k_x,k_y)}{\sqrt{\tau^2(k_x,k_y) N^{-1}(k_x,k_y)}},
\label{eqn:optfilt}
\end{equation}
where $\tau$ is the (Fourier-space, beam-and-filtering-convolved) source profile, 
and $N$ is the (Fourier-space) noise covariance matrix, which we assume to be diagonal.
The noise covariance can include contributions from unwanted sky signals (convolved 
with the beam and filtering) as well as 
instrumental and atmospheric noise.  For point sources, the source profile
is entirely a function of the beam and filtering:
\begin{equation}
\tau_\mathrm{PS}(k_x,k_y) = B(k_x,k_y) \ F(k_x,k_y) 
\label{eqn:taups}
\end{equation}
To extract source fluxes from the optimally filtered map, the values at the pixel 
location should be multiplied by the conversion factor in Table \ref{tab:bands} 
and an effective solid angle that now also includes the optimal filter $\psi$; 
however, if the filter is properly normalized as in Equation \ref{eqn:optfilt}, then
the effective solid angle will not change.

We can now estimate the flux of a known source in the SPT map using the raw
data and an optimally filtered map.  The radio
source PKS~0549-575 (also known as SPT-S~055009-5732.3) is located at 
R.A.~$87.5399^\circ$, decl.~$-57.5401^\circ$, which corresponds to pixel 
location [937,928] (zero-based indexing) in the oblique Lambert equal-area 
azimuthal projection maps.  The values in the point-source-unmasked 150 and 220~GHz
maps at this pixel location are $10.084 \times 10^{-3}$ and $8.508 \times 10^{-3}$
K-CMB.  Multiplying by the radio-source conversion factors in Table \ref{tab:bands}
yields values of $3.996$ and $4.056$~MJy/sr.  Multiplying the Fourier-space beam
by the filter transfer function appropriate to the source location
in each band and integrating over all $k$ values yields
effective solid angles of $1.523 \times 10^{-7}$ and $1.220 \times 10^{-7}$ sr, leading
to source flux estimates of $0.6086$ and $0.4949$ Jy.  We can estimate the statistical
uncertainty on these values by measuring the pixel variance in the neighborhood
of this source and converting that number to Jy.  This procedure yields rms uncertainties
of $0.0043$ and $0.0081$ Jy at 150 and 220~GHz.

\begin{figure*}[ht]
  \includegraphics[scale=0.68]{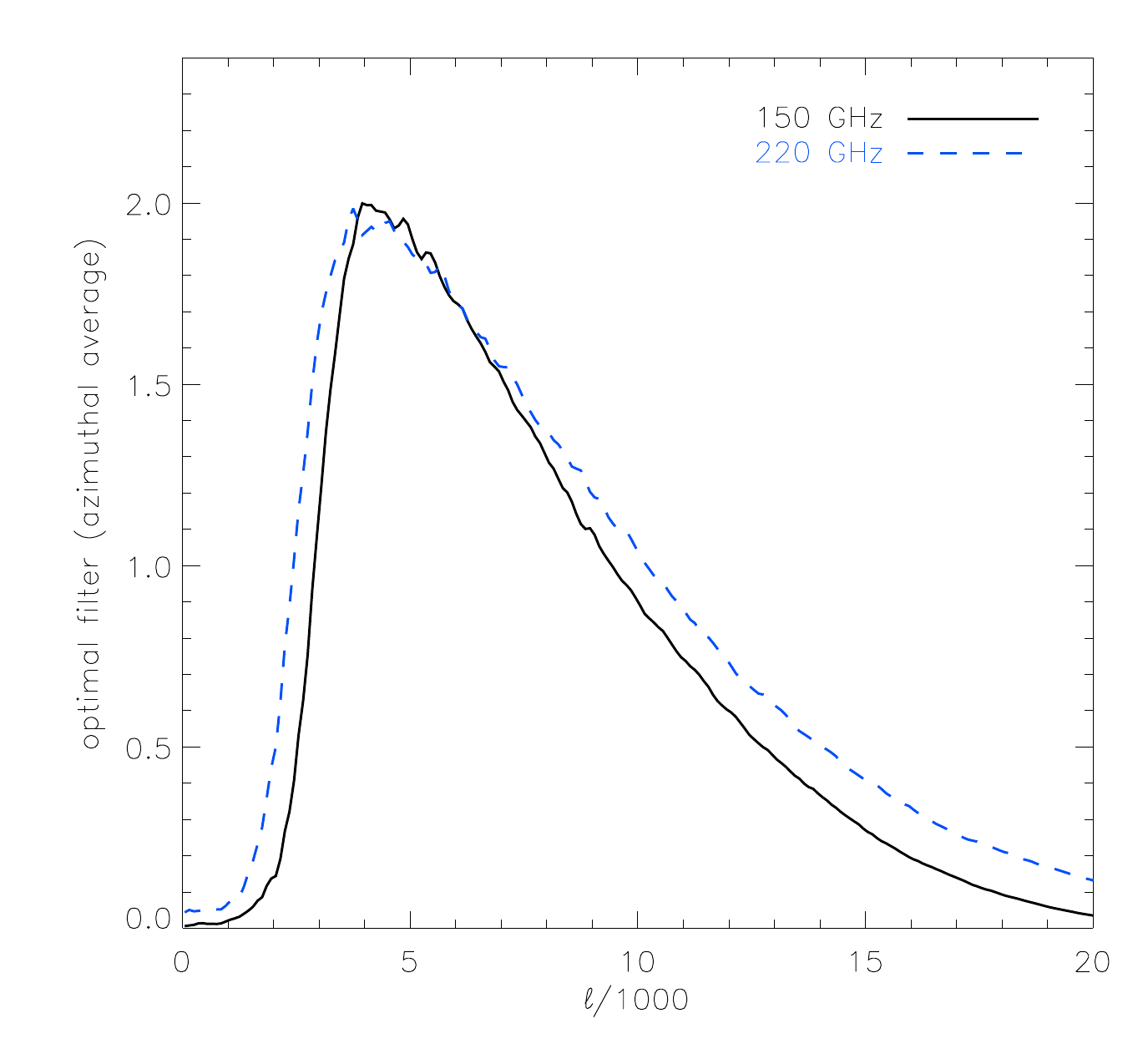}

\caption{Azimuthally averaged optimal filters for point-source extraction, constructed 
using the data products in this work and the CMB power spectrum from \citet{larson11}.  
\label{fig:optfilt} }
\end{figure*}

If we assume that the only signals in the map are CMB fluctuations and point sources, 
we can build an optimal filter using the source profile defined in Equation \ref{eqn:taups}
and a noise power spectrum that is the sum of the (squared) instrument-plus-atmosphere
PSD and the CMB power spectrum.
Because we have filtered the CMB, the version of
the power spectrum that goes into the optimal filter must be multiplied by 
$[B(k_x,k_y) F(k_x,k_y)]^2$.
Figure \ref{fig:optfilt} shows the azimuthally averaged optimal filter at 150 and 220~GHz
using the data products presented here and the CMB power spectrum from \citet{larson11}.  After 
convolving the maps with these filters, the map values at the location of PKS~0549-75
are $10.263 \times 10^{-3}$ and $8.761 \times 10^{-3}$ K-CMB at 150 and 220~GHz.  
As expected, the effective solid 
angle is the same as before application
of the optimal filter, and the source flux estimates obtained by multiplying these 
optimally filtered map values by the radio-source conversion factors in Table \ref{tab:bands}
and the solid angles calculated above are
$0.6194$ and $0.5096$ Jy.  Estimating the flux uncertainties as we did with the unfiltered map
yields rms uncertainties of $0.0016$ and $0.0032$ Jy at 150 and 220~GHz.  For a source
this bright, these statistical uncertainties are dwarfed by the calibration and beam
uncertainties.

\bibliography{../../../BIBTEX/spt.bib}

\end{document}